\def\spose#1{\hbox to 0pt{#1\hss}}

\def\multleft#1{\hbox to size{\vbox {\halign {\lft{##}\cr #1}}\hfill}\par}
\def\multright#1{\hbox to size{\vbox {\halign {\rt{##}\cr #1}}\hfill}\par}

\def\today{\ifcase\month\or January\or February\or March\or April\or May\or
      June\or July\or August\or September\or October\or November\or December\fi
      \space\number\day, \number\year}

\def\asec{$^{\prime\prime}$}

\def\H2{\hbox{H$_{2}$}}

\documentclass{mn2e}
\usepackage{times}
\usepackage{amssymb}
\usepackage{epsfig}
\usepackage{lscape}
\usepackage{graphicx}
\begin{document}
\hsize=6truein
          
\title[The $\bmath{z=7-9}$ galaxy luminosity function]
{A new multi-field determination of the galaxy luminosity function at $\bmath{z=7-9}$ incorporating
the 2012 Hubble Ultra Deep Field imaging}
\author[R.J.~McLure et al.]
{R. J. McLure$^{1}$\thanks{Email: rjm@roe.ac.uk}, J. S. Dunlop$^{1}$, R.A.A. Bowler$^{1}$, E. Curtis-Lake$^{1}$,
M. Schenker$^{2}$, \and R.S. Ellis$^{2}$, B.E. Robertson$^{3}$, A.\,M.~Koekemoer$^{4}$, A.B. Rogers$^{1}$,
Y. Ono$^{5}$, M. Ouchi$^{5}$, \and S. Charlot$^{6}$, V. Wild$^{7}$, D.P. Stark$^{3}$, S.R. Furlanetto$^{8}$,
M. Cirasuolo$^{1}$, T.A. Targett$^{1}$
\footnotesize\\
$^{1}$SUPA\thanks{Scottish Universities Physics Alliance}, Institute
for Astronomy, University of Edinburgh, Royal Observatory, Edinburgh
EH9 3HJ\\ 
$^{2}$Department of Astrophysics, California Institute of Technology, MS 249-17, Pasadena, CA 91125, USA\\
$^{3}$Department of Astronomy and Steward Observatory, University of Arizona, Tucson AZ 85721, USA\\
$^{4}$Space Telescope Science Institute, 3700 San Martin Drive, Baltimore, MD 21218, USA\\
$^{5}$Institute for Cosmic Ray Research, University of Tokyo, Kashiwa City, CHiba 277-8582, Japan\\
$^{6}$UPMC-CNRS, UMR7095, Institut d'Astrophysique de Paris, F-75014, Paris, France\\
$^{7}$School of Physics and Astronomy, University of St Andrews, North Haugh, St Andrews, KY16 9SS\\
$^{8}$ Department of Physics \& Astronomy, University of California, Los Angeles, CA 90095, USA}

\maketitle

\begin{abstract}
We present a new determination of the UV galaxy luminosity function (LF) at 
redshift $z \simeq 7$ and $z \simeq 8$, and a first estimate at $z\simeq 9$. An accurate determination of the form and evolution 
of the galaxy LF during this era is of key importance for improving 
our knowledge of the earliest phases of galaxy evolution and the process of cosmic reionization. Our analysis exploits 
to the full the new, deepest WFC3/IR imaging from our {\it Hubble 
Space Telescope} (HST) Ultra Deep Field 2012 (UDF12) campaign, with 
dynamic range provided by including a new and consistent analysis of all 
appropriate, shallower/wider-area HST survey data. 
Our new measurement of the evolving LF at $z \simeq 7 - 8$ 
is based on a final catalogue of $\simeq 600$ galaxies, and 
involves a step-wise maximum likelihood determination based on 
the photometric-redshift probability distribution for each object; 
this approach makes full use of the $11$-band imaging now available 
in the Hubble Ultra Deep Field (HUDF), 
including the new UDF12 F140W data, and the latest {\it Spitzer} 
IRAC imaging. The final result is a 
determination of the $z \simeq 7$ LF extending down to UV absolute magnitudes 
$M_{1500} = -16.75$ (AB mag), and the $z \simeq 8$ LF down to $M_{1500} = -17.00$. 
Fitting a Schechter function, we find $M_{1500}^{*} = -19.90^{+0.23}_{-0.28}$, 
$\log \phi^* = -2.96^{+0.18}_{-0.23}$, and a faint-end slope 
$\alpha = -1.90^{+0.14}_{-0.15}$ at 
$z \simeq 7$, and $M_{1500}^{*} = -20.12^{+0.37}_{-0.48}$, $\log \phi^* = -3.35^{+0.28}_{-0.47}$, 
$\alpha = -2.02^{+0.22}_{-0.23}$ at $z \simeq 8$. These results strengthen 
previous suggestions that the evolution at $z > 7$ appears more akin to 
`density evolution' than the apparent `luminosity evolution' seen at 
$z \simeq 5 - 7$. We also provide the first meaningful information on the 
LF at $z \simeq 9$, explore alternative extrapolations to higher redshifts, and 
consider the implications for the early evolution of UV luminosity density.
Finally, we provide catalogues (including derived $z_{phot}$, $M_{1500}$ and photometry) for the 100 most robust $z \simeq 6.5 - 11.9$
galaxies in the HUDF used in this
analysis. We briefly discuss our results in the context of earlier work 
and the results derived from an independent analysis of the UDF12 
data based on colour-colour
selection (Schenker et al. 2013).
\end{abstract}

\section{INTRODUCTION}
The advent of deep near-infrared imaging, in particular with Wide Field 
Camera 3 (WFC3/IR) on the {\it Hubble Space Telescope} (HST) has now enabled
the discovery and study of galaxies to be extended to redshifts 
$z \simeq 6.5 - 10$, into the first billion years of cosmic history
(see Dunlop 2012 for a review). This work is of fundamental importance 
for improving our understanding of the formation and growth of the early 
generations of galaxies, and testing the predictions of the latest 
galaxy-formation simulations. It is also of interest for establishing 
whether these galaxies reionized the Universe (e.g. Robertson et al. 
2010; Finkelstein et al. 2012; Kuhlen \& Faucher-Gigu\'{e}re 2012), 
and, if so, providing more detailed information on how reionization 
proceeded (as compared to the integrated constraints on `instantaneous' 
reionization provided by current measurements of microwave background 
polarization -- $z_{reion} \simeq 10.6 \pm 1.2$; Komatsu et al. 2011).

Because galaxies at $z \simeq 7$ are so faint, it is hard to gain  
detailed physical information on the properties of individual objects, and 
indeed only a handful of spectroscopic redshifts have 
been established on the basis of Lyman-$\alpha$ emission at $z \simeq 7$
(the current record holder is at $z = 7.213$; Ono et al. 2012\footnote{We note that the claimed
detection of a Lyman-$\alpha$ emitter at $z=8.55$ (Lehnert et al. 2010) now appears spurious (Bunker et al. 2013).}). Attention has
thus (sensibly) focussed on population statistics, helped by the fact that 
significant samples of photometrically-selected galaxies can now be 
assembled at these redshifts due to the presence of 
a strong Lyman-break at $\lambda_{rest} \simeq 1216$\,\AA\  
in their spectral energy distributions
(SEDs). This is caused by near-complete absorption by neutral 
hydrogen gas along the line-of-sight at $z > 6.5$ 
(Fan et al. 2006; Mortlock et al. 2011), making 
`Lyman-break' galaxy (LBG) selection in principle straightforward 
at these redshifts, given adequate data. 

The first and most important population measurement which is usually 
attempted once a significant sample of galaxies is available at
a given redshift is a determination of the luminosity function (LF); 
i.e. the comoving number density of galaxies as a function of 
luminosity ($\equiv$
absolute magnitude). Prior to the 2009 installation of WFC3/IR 
the availability of deep $z_{850}$ and $i_{775}$ imaging from the 
Advanced Camera for Surveys (ACS) on {\it HST} enabled the UV ($\lambda_{rest} 
\simeq 1500$\,\AA) LF to be established for faint galaxies out to 
$z \simeq 6$ by Bouwens et al. (2007). In a complementary effort based 
on the new availability of degree-scale red/infrared imaging from 
ground-based telescopes, the bright end of the UV LF was also measured 
out to $z \simeq 6$ by McLure et al. (2009), who demonstrated 
that a combined analysis yielded a consistent result. Both of these studies
exploited the available photometry to establish the presence/location 
of the afore-mentioned Lyman-break, but whereas Bouwens et al. (2007) 
continued with the simple and well-established two-colour selection technique, McLure et al. (2009)
used SED fitting with evolutonary synthesis models (e.g. 
Bruzual \& Charlot 2003) to derive photometric redshifts based on 
{\it all} of the available multi-band photometry. 

These two alternative approaches to galaxy selection have now 
both been exploited to explore the form of the LF at higher redshifts. 
Specifically, with the new first-epoch deep WFC3/IR data provided by the UDF09 
program (GO 11563, PI: Illingworth), Oesch et al. (2010a) and McLure et al.
(2010) produced alternative (but again consistent) determinations of the 
$z \simeq 7$ galaxy UV LF. The SED fitting approach was also exploited 
by Finkelstein et al. (2010), while colour-colour selection has since been 
re-applied by Bouwens et al. (2011a) at $z \simeq 7$ and $z \simeq 8$, to 
the final UDF09 dataset. Colour-colour selection has also recently 
been applied in attempts to constrain the brighter end of the LF at 
$z \simeq 8$ by Bradley et al. (2012) and Oesch et al. 
(2012b), to the BoRG\footnote{https://wolf359.colorado.edu/}  and 
CANDELS (Grogin et al. 2011)\footnote{http://candels.ucolick.org} datsets 
respectively. Meanwhile, SED fitting has been applied to the UDF09 and 
CANDELS GOODS-South data by Finkelstein et al. (2012) in order to derive 
a new estimate of the evolving UV luminosity density, and by Bowler et al. (2012) in the 
search for brighter $z \simeq 7$ galaxies in the early UltraVISTA data 
(McCracken et al. 2012)\footnote{http://www.ultravista.org}.

It is important to note that, given only three-filter data (i.e. two-wavebands
above a putative Lyman-break and one below) colour-colour selection and 
SED fitting are essentially equivalent. However, as the number of useful
wavebands expands, it is clear that SED fitting makes more complete and 
consistent use of the available data. This has, to some extent,  
been recognized by the adoption of additional criteria 
to colour-colour selection,
in an attempt to `factor-in' the extra information provided through
other filters (e.g. the rejection of objects which which show more 
that one $> 1.5$-$\sigma$ detection in bluer bands and the computation 
of a separate $\chi^2_{optical}$ by Bouwens et al. 2011a).
However, SED fitting clearly deals with all detections and non-detections in 
a more straightforward and consistent manner, and has the additional benefit 
of providing actual redshift estimates with 
confidence intervals (and indeed can provide a redshift probability
distribution for each object, albeit this depends somewhat on adopted
priors; McLure et al. 2011). Finally, SED fitting also more clearly exposes the nature of 
potential interlopers (such as dusty red galaxies, post-starburst 
objects with strong Balmer breaks, and dwarf stars in our own galaxy) and 
provides clearer information on what data needs to be improved to eliminate them 
(e.g. Dunlop 2012). Nevertheless, SED fitting and colour-colour selection both fundamentally rely on the Lyman-break and, whatever the selection technique, careful simulation 
work is required to quantify selection bias, completeness and contamination 
in any determination of the evolving galaxy LF in the young universe.

In an attempt to make further progress, and in particular to extend the study of galaxies
both to higher redshifts ($z > 8$) and lower luminosities (at $z \simeq 7 - 8$) we have 
recently completed a new programme of even deeper near-infrared imaging in 
the Hubble Ultra Deep Field (HUDF; Beckwith et al. 2006) with WFC3/IR on {\it HST}.
This new imaging campaign was completed in September 2012 (GO 12498, PI: Ellis, hereafter 
UDF12) and when combined with the existing UDF09 data provides the deepest ever 
near-infrared images of the sky. The data have now been reduced and released to the 
public (Koekemoer et al. 2013) through the team website\footnote{http://udf12.arizona.edu}.
Key elements of our observing strategy were the delivery of extremely deep $Y_{105}$ 
data to more robustly identify galaxies at $z \simeq 8$ and higher redshifts, and the addition
of imaging through a new filter previously unexploited in the HUDF, $J_{140}$,  
both to enable reliable galaxy discovery to be pushed beyond $z \simeq 8.5$ (with two filters long-ward of the Lyman break), and to enable more accurate 
SEDs to be determined for galaxies at $z \simeq 7$ and $z \simeq 8$. The final UDF12+UDF09
combined dataset reaches the planned 5-$\sigma$ detection limits of $Y_{105} = 30.0$, $J_{125} = 29.5$,
$J_{140} = 29.5$, $H_{160} = 29.5$ (in apertures of diameter 0.40\asec, 0.44\asec, 0.47\asec, 0.50\asec  
respectively, sampling 70\% of point-source flux density in each waveband). The results of 
our search for galaxies at $z > 8.5$ have already been reported by Ellis et al. (2013), while 
the new deep multi-band data have now also been exploited by Dunlop et al. (2013) in a new
determination of the UV spectral slopes of galaxies at $z \simeq 7 - 9$ (with consequent 
implications for their stellar populations). Most recently, a new determination of galaxy 
sizes based on the UDF12 dataset has been completed by Ono et al. (2013).

In this paper we focus on utilising the UDF12 dataset, along with the ever-growing shallower 
WFC3/IR imaging over wider areas, to undertake a new determination of the galaxy UV LF at $z \simeq 7$ 
and $z \simeq 8$. Crucially the new ultra-deep imaging in the HUDF improves our ability to probe the faint end of 
the LF, better sampling the population of numerous faint galaxies ($M_{1500} > -18$) which likely 
dominate the UV luminosity density and hence drive reionization. A key goal, therefore, is to better 
establish the faint-end slope, $\alpha$, on which extrapolations to even fainter (as yet unobservable)
luminosities have to be based. However, simply increasing the depth of the deepest field does 
not yield significantly better estimates of $\alpha$ unless the degeneracies between the Schechter function 
parameters ($M^*$, $\phi^*$, and $\alpha$) can be minimized (e.g. Bouwens et al. 2011a; Dunlop 2012). 
This requires maximising the usable 
dynamic range in UV luminosity, to properly constrain the shape of the LF. Thus, 
to best determine the $z \simeq 7$ and $z \simeq 8$ LF, 
we have analysed the new HUDF12 data in combination with the progressively shallower 
WFC3/IR survey data provided by the UDF09 parallel fields, the Early Release 
Science (ERS) 
data in GOODS-South, the CANDELS data in the remainder of GOODS-South and the UDS field, 
and all of the parallel BoRG data obtained by Sept 2012. In each of these fields, 
(including the HUDF) we have also utilised the  associated HST ACS imaging, 
{\it Spitzer} IRAC data (deconfused with the WFC3/IR $H_{160}$ imaging), and 
ground-based near-infrared/optical data where appropriate. 

Given that the selection functions and associated simulations are different, and to facilitate comparison with other 
work, our team has also undertaken a parallel, and completely independent determination 
of the LF at $z \simeq 7$ and $z \simeq 8$ based on `traditional' drop-out colour-colour selection.
The results from this are presented in Schenker et al. (2013) but are also summarised in this 
present paper for ease of comparison with the SED-fitting technique results derived here. 

The remainder of this paper is structured as follows. In Section 2 we give 
full details of the datasets utilised in this new study, and explain how we selected 
galaxy catalogues before refining the samples to contain only plausible high-redshift galaxy 
candidates. We also describe the simulations undertaken to establish completeness and 
contamination corrections in each of the individual survey fields, simulations which are crucial 
for a robust determination of the LF from such a complex multi-field dataset. Next, in Section 3, 
we describe how we chose to determine the LF, adopting as our primary technique 
the non-parametric step-wise maximum likelihood method (SWML), but also applying 
parametric maximum likelihood fitting to explore Schechter-function representations of the LF. We then present the results of our analysis in Section 4,
providing our best measurements of the LF at $z \simeq 7$, $z \simeq 8$ and $z \simeq 9$, and briefly 
exploring the implied evolution of the LF with redshift. Here we also compare our results 
with the independent UDF12 analysis of Schenker et al. (2013), and discuss our 
derived LF parameters (with associated improved confidence intervals) in the context 
of the the results deduced 
by Bouwens et al. (2011a), Bradley et al. (2012) and Oesch et al. (2012b) prior to UDF12.
In Section 5 we proceed to explore the implications of our results for the evolution of the LF 
out to even higher redshifts, and derive the implied evolution of UV luminosity density 
as a function of redshift (a key measurement for tracking the likely progess 
of reionization). Finally, we present a summary of our conclusions
in Section 6. Throughout the paper we will refer to the following {\it HST} ACS+WFC3/IR filters: 
F435W, F600LP, F606W, F775W, F814W, F850LP, F098M, F105W, F125W, F140W \& F160W as 
$B_{435}, V_{600}, V_{606}, i_{775}, i_{814}, z_{850}, Y_{098}, Y_{105}, J_{125}, J_{140}\, \&\,H_{160}$ respectively.
All magnitudes are quoted in the
AB system (Oke 1974; 
Oke \& Gunn 1983) and all cosmological 
calculations assume $\Omega_{0}=0.3, \Omega_{\Lambda}=0.7$ and $H_{0}=70$ kms$^{-1}$Mpc$^{-1}$. 

\section{Data}
In this section we provide a summary of the basic properties of the datasets which we have utilised in this study to 
measure the high-redshift galaxy luminosity function. In addition, we also provide details of the methods
adopted to derive accurate image depth information, object photometry and reliable catalogues of high-redshift galaxy candidates.

\begin{table*}
\caption{The basic observational properties of the different WFC3/IR survey fields analysed in this work. 
Column 1 lists the adopted field name, columns 2 \& 3 list the coordinates of the centre of the field and column 4 lists the 
survey area in arcmin$^2$. Columns 5-14 list the global average $5\sigma$ depths, which have been corrected to total magnitudes 
assuming a point source (a typical correction of $\simeq 0.2$ magnitudes for ACS and $\simeq 0.4$ magnitudes for WFC3/IR; see Section 2.3 for a 
discussion of the apertures adopted in each filter). It should be noted that the depths within a given field can vary 
significantly (e.g. CANDELS GS DEEP and BoRG). There are no coordinates listed for BoRG simply because it consists of a large
number of widely-separated pointings.}
\begin{tabular}{lccrccccccccccc}
\hline
Field & RA(J2000)&Dec(J2000)      &Area           &$B_{435}$&$V_{606}$&$i_{775}$&$i_{814}$&$z_{850}$&$Y_{098}$&$Y_{105}$&$J_{125}$&$J_{140}$&$H_{160}$\\
\hline
HUDF             &03:32:38.5&$-27$:46:57.0&  4.6  & 29.7   & 30.2   & 29.9  & $-$    &29.1   & $-$     & 29.7  & 29.2  & 29.2 & 29.2\\ 
HUDF09-1         &03:33:01.4&$-27$:41:11.5&  4.4  & $-$    & 29.1   & 28.9  & $-$    &28.6   & $-$     & 28.5  & 28.6  & $-$  & 28.2\\ 
HUDF09-2         &03:33:05.5&$-27$:51:21.6&  4.5  & $-$    & 29.2   & 29.0  & 29.8   &28.6   & $-$     & 28.6  & 28.7  & $-$  & 28.3\\ 
CANDELS GS-WIDE  &03:32:38.5&$-27$:53:36.5& 35.1  & 28.0   & 28.4   & 27.8  & $-$    &27.5   & $-$     & 26.9  & 27.1  & $-$  & 26.6\\
ERS              &03:32:23.4&$-27$:42:52.0& 38.4  & 28.0   & 28.4   & 27.8  & $-$    &27.5   & 27.2    & $-$   & 27.4  & $-$  & 27.0\\
CANDELS GS-DEEP  &03:32:29.8&$-27$:47:43.0& 64.6  & 28.0   & 28.4   & 27.8  & $-$    &27.5   & $-$     & 27.9  & 27.7  & $-$  & 27.3\\
CANDELS UDS      &02:17:25.7&$-05$:12:04.6& 144.5 & $-$    & 27.6   & $-$   & 27.5   & $-$   & $-$     & $-$   & 26.8  & $-$  & 26.7\\
BoRG             &$-$       &$-$          & 180.4 & $-$    & 27.5   & $-$   & $-$   & $-$   & 27.8     & $-$   & 26.8  & $-$  & 26.8\\
\hline
\end{tabular}
\end{table*}

\subsection{Survey fields}
The datasets analysed in this paper form a `wedding-cake' structure ranging from the ultra-deep UDF12 observations covering
an area of only $\simeq 4.5$ arcmin$^2$ to wider-area WFC3/IR survey data covering several hundred arcmin$^2$. Below we provide 
the basic observational details of each dataset in turn.

\subsubsection{The UDF12}
The dataset which plays the pivotal role in constraining the faint-end of the galaxy luminosity
function at $z\geq 7$ and provides the primary motivation for this paper is the new UDF12 WFC3/IR multi-band imaging
of the Hubble Ultra Deep Field (GO-12498, P.I. Ellis). The UDF12 observing campaign acquired 128 orbits of WFC3/IR integration 
time targeting the HUDF, all of which were obtained between 4th August and 16th September
2012\footnote{The entire reduced UDF12 dataset is publicly available on the following
web-site: http://archive.stsci.edu/prepds/hudf12/}. As discussed in the introduction, the primary motivation
for the UDF12 observing campaign was to improve our knowledge of number
densities and spectral properties of the ultra-faint galaxy
population at $z=7-8$ and to provide the first robust census of the
$z\geq 8.5$ galaxy population. 

In order to achieve these aims the
bulk of the UDF12 orbits were invested in quadrupling the HUDF integration time in the crucial $Y_{105}$ 
filter and providing ultra-deep imaging in the $J_{140}$ filter, 
which had not been
employed in previous HUDF imaging campaigns. The 128 orbits awarded to UDF12 were
allocated as follows: 72 orbits in the $Y_{105}$ filter, 30 orbits in
the $J_{140}$ filter and 26 orbits in $H_{160}$. 
In combination with the data provided by the previous UDF09 observing
campaign (GO-11563, P.I. Illingworth), the total orbit allocation of
dedicated WFC3/IR data in the HUDF now stands at: 96 orbits in $Y_{105}$, 34
orbits in $J_{125}$, 30 orbits in $J_{140}$ and 79 orbits in
$H_{160}$. The depths of the available data in the HUDF (and the other survey fields analysed in this study) 
are provided in Table 1. It should be noted that the depths quoted in Table 1 have been corrected to total magnitudes assuming a point-source
and that the raw aperture depths are $0.2-0.4$ magnitudes deeper, depending on the adopted aperture (see Section 2.3).

\subsubsection{The HUDF09 parallel fields}
To increase our ability to constrain the faint-end of the high-redshift luminosity function we have also utilised the WFC3/IR 
imaging available in the two parallel fields of the HUDF09 imaging campaign, hereafter HUDF09-1 and HUDF09-2. Although substantially 
shallower than the data available in the HUDF itself, the HUDF09 parallel fields consist of 33 and 48 WFC3/IR orbits 
respectively (spread between the $Y_{105}, J_{125}\, \&\, H_{160}$ filters) and provide crucial leverage for constraining the $z\geq 7$ 
galaxy luminosity function $\simeq 1.5$ magnitudes brighter than the ultimate depth achieved by UDF12.
For the purposes of this analysis we have utilised our own reduction of the HUDF09 WFC3/IR imaging in both parallel 
fields (drizzled onto a 30mas pixel scale) and have also used our own reduction of the ACS imaging covering HUDF09-2 originally
obtained as part of the HUDF05 campaign (GO-10632, P.I. Stiavelli). For the HUDF and HUDF09-1 we have made use of the 
publicly-available reductions of the original HUDF ACS imaging (Beckwith et al. 2006) and the HUDF05 ACS imaging respectively. Finally, we have also made use of the new ultra-deep $i_{814}$ ACS imaging (128 orbits) obtained as parallel 
observations during the UDF12 campaign which provides a $\simeq60\%$ 
overlap with HUDF09-2.

\subsubsection{GOODS-S}
In addition to the HUDF and parallel fields, we have made extensive use of the publicly-available WFC3/IR imaging of the 
Great Observatories Origins Deep Survey South (GOODS-S) field provided by the Cosmic Assembly Near-Infrared Deep Extragalactic Legacy Survey (CANDELS, Grogin et al. 2011; Koekemoer et al. 2011). The CANDELS WFC3/IR data in 
GOODS-S covers a total of 24 WFC3/IR pointings which are divided into DEEP and WIDE sub-regions. The DEEP sub-region consists 
of 15 WFC3/IR pointings each consisting of 3 orbits of $Y_{105}$ imaging and 5 orbits in both $J_{125}$ and $H_{160}$. 
The WIDE sub-region consists of 9 WFC3/IR pointings each consisting of a single orbit of integration in the $Y_{105}$, $J_{125}$ and 
$H_{160}$ filters. In addition to the CANDELS imaging we have also 
analysed the ERS data in GOODS-S which consists of 
10 WFC3/IR pointings, each of 
which were observed for 2 orbits in the $Y_{098}$, $J_{125}$ and $H_{160}$ filters (Windhorst et al. 2011).

The optical data we have employed in GOODS-S is the publicly-available v2.0 reduction of the original GOOD-S ACS imaging 
in the $B_{435}, V_{606}, i_{775} \,\&\,z_{850}$ filters (Giavalisco et al. 2004). The total area of the overlapping WFC3/IR+ACS 
coverage we have analysed in GOODS-S is 138 arcmin$^2$ (excluding the HUDF and parallel fields). A summary 
of the available filters and depths is provided in Table 1.

\subsubsection{The UKIDSS Ultra Deep Survey Field}
In order to significantly increase the total areal coverage of our dataset, crucial for constraining the bright end of the 
$z\geq 7$ galaxy luminosity function, we have also analysed the publicly-available CANDELS WFC3/IR+ACS imaging in the UKIDSS 
Ultra Deep Survey (UDS; Lawrence et al. 2007). The full CANDELS dataset in the UDS consists of 44 WFC3/IR pointings, each featuring 4/3 of an orbit 
of $H_{160}$ imaging and 2/3 of an orbit in $J_{125}$. Along with the primary WFC3/IR observations, the CANDELS campaign in the UDS 
also obtained ACS parallels with each pixel typically receiving three orbits of integration in $i_{814}$ and one orbit in $V_{606}$. 
However, due to the focal plane separation of the WFC3/IR and ACS cameras, only 32/44 of the WFC3/IR pointings are fully covered by 
the parallel ACS imaging. It is from these 32 WFC3/IR pointings, covering a total area of $\simeq 150$\,arcmin$^2$, that we have 
selected our sample of high-redshift candidates in the UDS.

In addition to the primary {\it HST} imaging data, in refining our high-redshift 
sample in the UDS we have also made extensive use of a variety of ground-based datasets. 
Amongst these the three most important are new ultra-deep $z^{\prime}-$band 
imaging obtained with Suprime-Cam on Subaru which reaches a depth of $z^{\prime}=26.5$ ($5\sigma$; 1.8\asec diameter apertures), 
new ultra-deep VLT+HAWK-I $Y-$band observations of the UDS CANDELS region obtained as part of the HUGS programme (P.I. A. Fontana) 
which reach a depth of $Y=26.5$ ($5\sigma$; 1.25\asec diameter apertures) and the latest DR10 release of the UDS $K-$band imaging 
which reaches a depth of $K=25.1$ ($5\sigma$; 1.8\asec diameter apertures). Moreover, in order to clean the sample of low-redshift 
interlopers, we have utilised a stack of the $BVRi^{\prime}$ Subaru imaging of the UDS described in Furusawa et al. (2008), which 
reaches a depth of $\geq 29$ ($2\sigma$; 1.8\asec diameter apertures). Finally, in order to further refine our photometric redshift 
solutions we have exploited narrow-band (NB921, $\lambda_{C}=9210$\AA) imaging of the UDS (Sobral et al. 2011) which reaches a depth 
of $z_{921}=26.0$ ($5\sigma$; 1.8\asec diameter apertures). 

\subsubsection{BoRG}
In order to increase our ability to constrain the bright-end of the $z=8$ luminosity function we 
have performed our own reduction and analysis (Bowler et al. 2013, in preparation) of the data taken by the Brightest of the Reionizing Galaxies survey
(BoRG; Trenti et al. 2011, 2012).  BoRG is a HST pure-parallel
programme, consisting of imaging in four filters from WFC3, designed to
detect $z \sim 8$ Lyman-break galaxies as $Y_{098}$ drop-outs.
Details of the BoRG observation strategy can be found in Trenti
et. al. (2011), however briefly, pure-parallel observations were
obtained at multiple sightlines in the $Y_{098}$, $J_{125}$, $H_{160}$
filters from WFC3/IR and one or both of $V_{606}$ and $V_{600}$ with WFC3/UVIS.  The
exposure times are chosen to allow the detection of
$Y_{098}$ drop-out galaxies at $z \sim 8$ based on a large $Y_{098} - J_{125}$ colour and 
relatively flat rest-frame spectral slope inferred from the $J_{125} - H_{160}$ colour. 

At the time of writing the complete BoRG dataset consists of 69 independent fields with a variety of different exposure times. 
However, to homogenize the dataset somewhat, we have restricted our analysis to the 41 BoRG fields with $5\sigma$ detection limits 
in the range $J_{125}\geq 27.2 - 27.9$ (0.44\asec diameter aperture), and which lie at high galactic latitude. 
The calibrated FLT files were obtained from the HST archive and background subtracted 
before being combined with {\sc astrodrizzle} (Gonzaga et al. 2012). The final pixel size was set to 80mas to match the BoRG09 public reductions, using 
a large pix\_frac (pix\_frac=0.9 for multiple exposures and pix\_frac = 1.0 for single exposures) to account for the lack of dithering in the 
observations where the primary was often spectroscopic. 

Based on these 41 fields (total area 180\,arcmin$^2$) an initial candidate list was constructed using the colour-cuts employed by 
Bradley et al. (2012), but based on our own photometry and depth analysis (see below). As with all the other survey fields analysed here, these candidates 
were then analysed with our photometric redshift code before being included in the luminosity function analysis. To mitigate the increased photometric 
redshift uncertainties introduced by the small number of available filters, in our final analysis of the $z=8$ luminosity function we include only those 
candidates selected from BoRG brighter than $M_{1500}\leq -20.5$.

\begin{figure}
\includegraphics[width=8.0cm, angle=0]{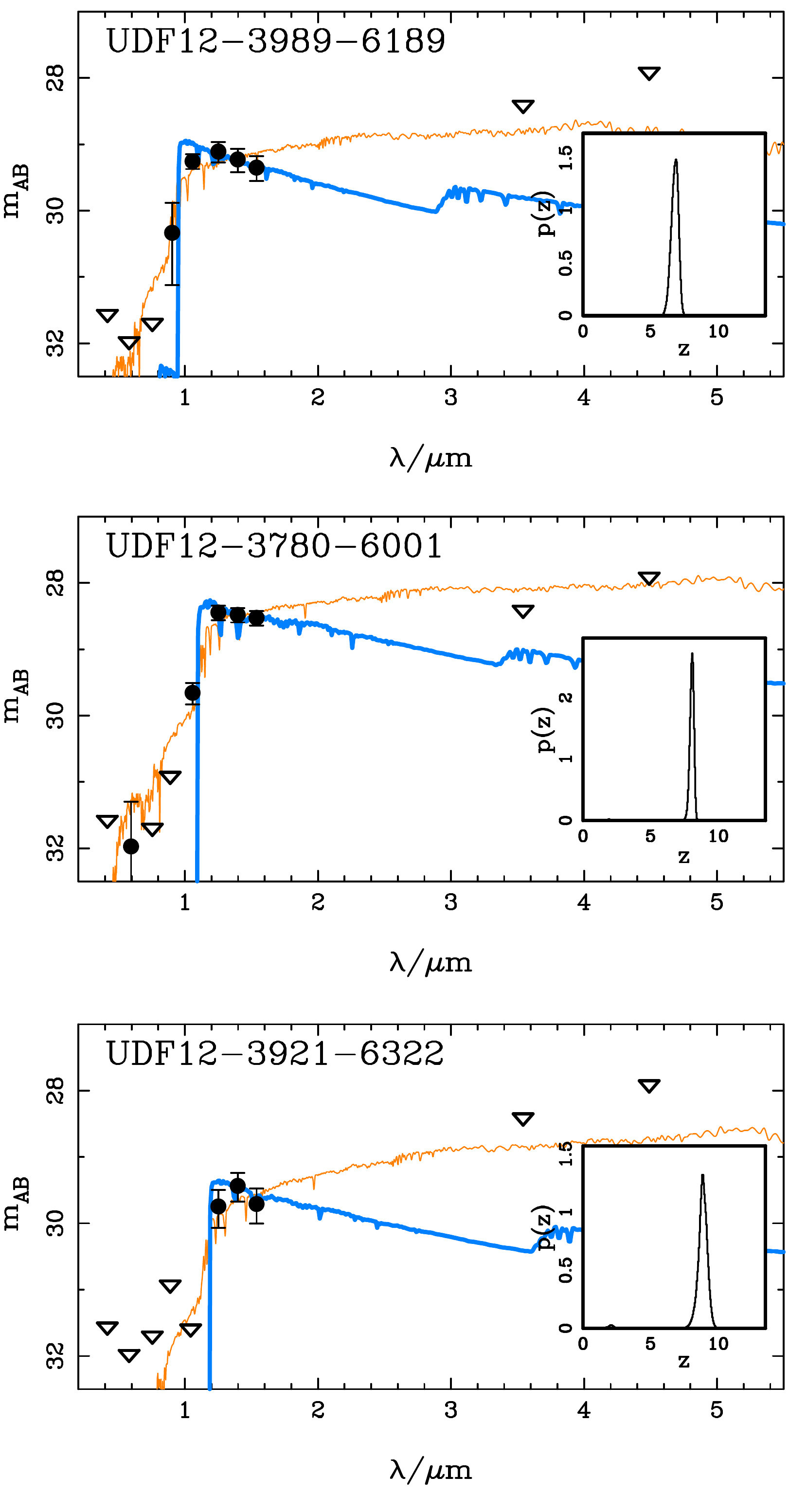}
\caption{Examples of our photometric redshift analysis for three objects in the HUDF at redshifts $z=6.9, z=8.1$ and $z=8.8$ (top to bottom).
In each plot the blue curve shows the best-fitting high-redshift solution, while the orange curve shows the best-fitting alternative
low-redshift solution. In each case the inset shows the photometric redshift probability density function, $p(z)$, which is 
incorporated into our luminosity-function analysis. The upper limits at $3.6\mu$m and $4.5\mu$m have been derived via our own 
deconfusion analysis of the ultra-deep IRAC imaging obtained by Labb\'{e} et al. (2012), using the technique 
described in McLure et al. (2011). All upper limits are $1\sigma$.}
\end{figure}

\subsection{Catalogue production}
Given that the primary focus of this paper is the evolution of the galaxy luminosity function at $z\geq 6.5$, by which point the Lyman-break 
has been redshifted to an observed wavelength $\lambda_{obs} \geq 0.9\mu$m, high-redshift galaxy candidate selection was performed exclusively 
in the near-IR using the WFC3/IR imaging available in each field.

In order to provide a master catalogue which was as complete as possible, objects were initially selected using each individual 
near-IR image {\it and} from every possible wavelength-contiguous stack of near-IR images. For example, based on the UDF12 data,
four object catalogues were generated with {\sc sextractor} v2.8.6 (Bertin \& Arnouts 1996) using the 
individual $Y_{105}$, $J_{125}$, $J_{140}$ and $H_{160}$ images for object detection and a futher six catalogues were generated 
using the following near-IR stacks as the detection image: $Y_{105}+J_{125}+J_{140}+H_{160}$, $Y_{105}+J_{125}+J_{140}$, $Y_{105}+J_{125}$, 
$J_{125}+J_{140}$, $J_{125}+J_{140}+H_{160}$ \& $J_{140}+H_{160}$. 

From this initial set of 10 object catalogues, a master catalogue was constructed containing every {\it unique} object which was detected 
at $\geq 5\sigma$ significance in any of the detection images. For those objects which were present in multiple catalogues, the positional 
information and photometry based on the highest signal-to-noise detection was propagated to the 
master catalogue. Although this specific example is within the context of the UDF12 dataset in the HUDF, the general selection 
process was identical in each field, notwithstanding differences enforced by the number of available filters.

\subsection{Photometry}
When selecting samples of high-redshift galaxies, the choice of photometric apertures is inevitably a balance
between optimizing depth and ensuring that it is possible to derive well-defined and stable aperture corrections.
The photometry adopted in this study is all based on circular apertures, where the choice of aperture in each band is 
tuned to enclose $\geq 70\%$ of the flux of the filter-specific point-spread function (PSF). For those fields with data drizzled 
onto a 0.03\asec/pix grid, the photometry is based on 0.3\asec diameter apertures in the optical ACS bands and 0.40\asec, 0.44\asec, 0.47\asec 
and 0.50\asec diameter apertures in the $Y_{105}$, $J_{125}$, $J_{140}$ and $H_{160}$ bands. 
For those datasets where the data is drizzled onto a 0.06\asec/pix grid, we have adopted 0.40\asec diameter apertures in the optical 
ACS and $Y_{098}$/$Y_{105}$ WFC3/IR bands, together with 0.44\asec and 0.50\asec diameter apertures in the $J_{125}$ and $H_{160}$ bands. 
For the purposes of the subsequent photometric redshift analysis (see below), the measured photometry in $z_{850}$ and the WFC3/IR filters 
was corrected to the same enclosed flux level as the $B_{435}, V_{606}, i_{775}\, \& \, i_{814}$ photometry (typically $82\%-84\%$) using the curve of growth of the observed PSFs in each band/field.

\subsection{Depth analysis}
A crucial element of reliable high-redshift candidate selection is the derivation of accurate information regarding the photometric depth of each
available image. This information is clearly vital for the reliable exclusion of low-redshift contaminants, but is also required in order to provide 
the robust flux measurement errors necessary for accurate photometric redshift results.

For each survey field in turn, accurate maps of which pixels contained significant object flux were constructed by 
stacking {\sc sextractor} segmentation 
maps for each available filter. Before stacking, the individual segmentation maps were dilated in order to better capture the extended wings 
of bright low-redshift objects. Based on these image maps, a large number of photometric apertures (typically 10-200 thousand) were located 
amongst the fraction of each image that had been determined to be dominated by `sky'. A robust estimator was then used to measure the sigma 
of the distribution of the fluxes measured within these sky apertures and thereby determine the empirical depth of the image for a given 
aperture diameter. This aperture-to-aperture r.m.s. depth measurement is a robust method of determining the actual significance of any aperture 
flux measurement and captures the true noise properties of a given image, which are typically underestimated by a simple measurement of the 
pixel r.m.s. due to small-scale noise correlations introduced by the reduction procedure (Koekemoer et al. 2013).
Once the true depth of each image was determined in this fashion, the corresponding weight maps were scaled to match the empirical depth measurement. By including the scaled weight maps in the dual-mode catalogue production process, it was thereby possible to provide accurate, position dependent, flux measurement errors for each source.

\begin{figure}
\includegraphics[width=8.0cm, angle=0]{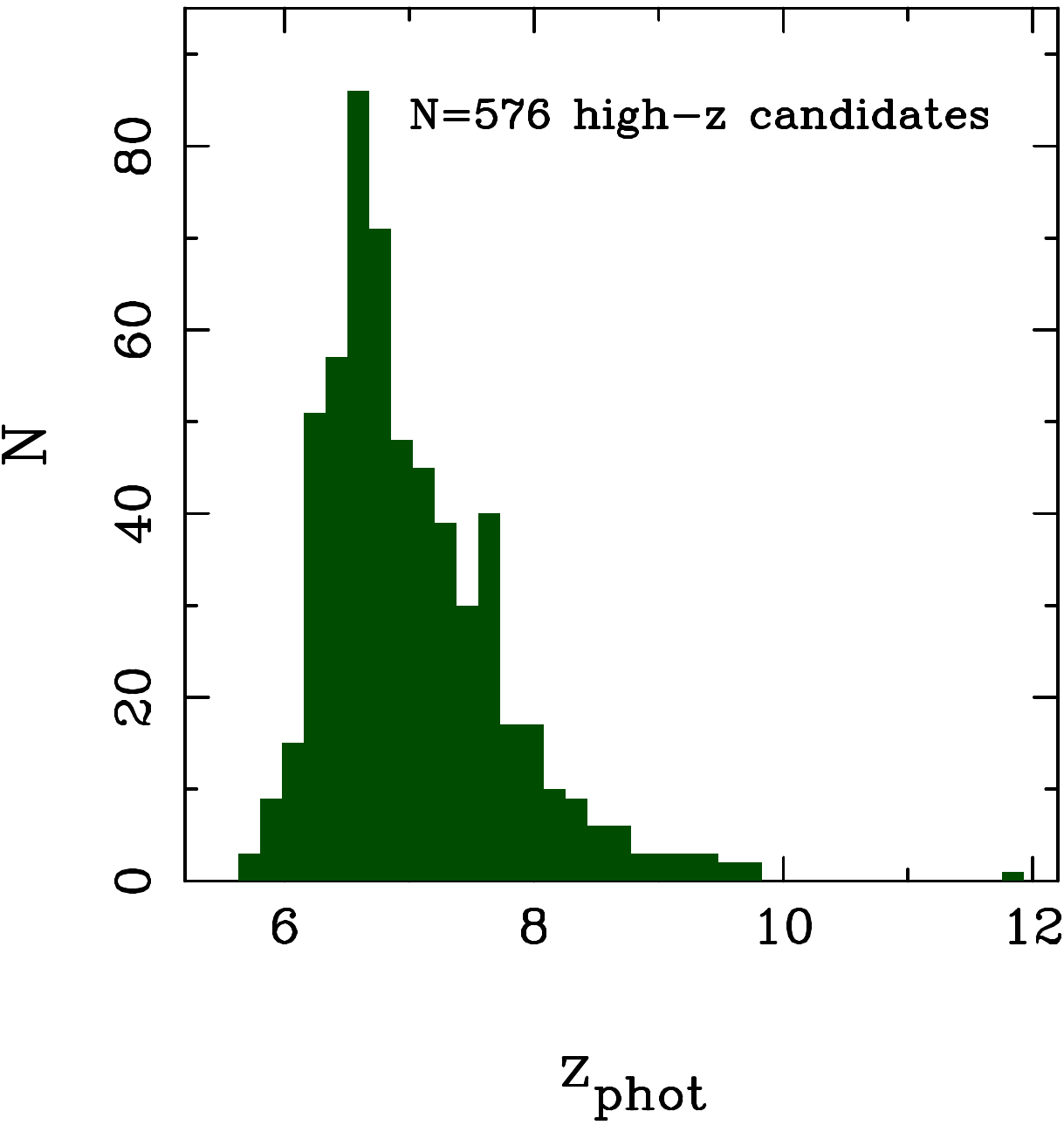}
\caption{The photometric redshift distribution of the full sample of high-redshift galaxy candidates ({\it robust}+{\it insecure}) derived
from our analysis of the eight survey fields listed in Table 1. The full sample consists of $N=576$ high-redshift candidates 
selected from a total area of 477 arcmin$^2$.}
\end{figure}

\subsection{High-redshift galaxy selection}
In order to study the $z\geq 7$ luminosity function it is necessary to derive reliable catalogues of $z\geq 6.5$ galaxy candidates.
Consequently, the master object catalogues for each survey field were initially cleaned of low-redshift contaminants by insisting 
that each object remained undetected at the $2\sigma$ level in each filter short-ward of the $z_{850}$ waveband. 
In addition, to exclude the small number of objects with $\simeq 1.5-2\sigma$ detections in multiple 
blue optical filters, each high-redshift candidate was also required to be undetected at the $2\sigma$ level in an inverse-variance weighted 
stack of all filters short-ward of $z_{850}$.

\subsubsection{Photometric redshift analysis}
After the initial exclusion of low-redshift contaminants, the master catalogues for each field were processed using our photometric 
redshift code. For a full description of the photometric redshift code the reader is referred to McLure et al. (2011), but we provide a 
brief outline here for completeness. The photometry for each galaxy is fitted with a range of galaxy templates, either empirical 
spectra or evolving synthetic galaxy-evolution models, with the best-fitting galaxy parameters determined via $\chi^{2}$-minimization. 
To ensure a proper treatment of the photometric uncertainties, the model fitting is performed in flux-wavelength space, rather than 
magnitude-wavelength space, and negative fluxes are included.

The code fits a wide range of reddening, based on the 
Calzetti et al. (2000) dust attenuation law, and accounts for IGM absorption according to the Madau (1995) prescription. If 
necessary, the code can also fit each high-redshift candidate including additional Ly$\alpha$ emission within a plausible 
range of rest-frame equivalent widths (chosen to be $EW_0\leq 240$\AA; Charlot \& Fall 1993). The best-fitting photometric redshifts and redshift probability density 
functions adopted here are based on Bruzual \& Charlot (2003) stellar population models with metallicities of 
0.2$Z_{\odot}$ or $Z_{\odot}$, although the exact choice of stellar population model has little impact on the derived $p(z)$ in most cases.
Three examples of the results of our photometric-redshift analysis are shown in Fig. 1; the three objects illustrated 
have derived redshifts of $z \simeq 6.9$, $z \simeq 8.1$ and $z \simeq 8.8$ (see Table A1).

Based on the photometric-redshift results, objects were excluded if it was impossible to obtain a statistically acceptable 
solution at $z\geq 6.5$ (typically $\chi^{2}_{best}\geq 20$), or if the photometric redshift probability density function indicated a 
very low probability that the object is at $z\geq 6.5$ (i.e. $\int_{z=6.5}^{z=\infty}p(z)dz \leq 0.05$). 

At this point, the catalogues were visually expected in order to remove spurious contaminants such as artefacts, diffraction spikes and 
over-deblended low-z objects. Following this final cleaning step, and following the procedure of McLure et al. (2011), each object was 
classified as {\it robust} or {\it insecure} depending on whether or not the best-fitting low-redshift solution could be ruled out at 
the $2\sigma$ confidence level. The objects classified as {\it robust} constitute our most reliable high-redshift galaxy sample and are 
presented in the Appendix for comparison with the results of other similar studies and potential spectrocopic follow-up observations. 
However, it should be noted that all high-redshift candidates 
which survived the full selection process, whether {\it robust} or 
{\it insecure}, were included in the luminosity function analysis (see below).

\subsubsection{Dwarf star exclusion}
As has been widely discussed in the literature (e.g. Dunlop 2012; Bowler et al. 2012), 
ultra-cool galactic dwarf stars (M, L, \& T dwarfs) are a potential source of low-redshift 
contamination of the bright-end of the galaxy luminosity function at $z > 6$. 
Within the context of the current 
study, the most serious source of concern is potential contamination of the $z=7$ galaxy luminosity function by T-dwarf stars.
To deal with this problem the photometry of all bright $z\simeq 7$ LBG candidates was analysed using a 
library of empirical optical-to--near-IR dwarf star spectra spanning the spectral range M0 to T9, taken 
from the SpeX archive\footnote{http://pono.ucsd.edu/~adam/browndwarfs/spexprism/}. All bright candidates which returned a high-quality fit with a dwarf star 
template and had a measured WFC3/IR half-light radius which was consistent with being spatially unresolved were removed from the final 
sample.

\subsection{Final galaxy sample}
The final high-redshift galaxy sample utilised in this analysis comprises 
a total of $N=576$ galaxy candidates ({\it robust}+{\it insecure})
selected from a total survey area of 477\,arcmin$^2$.
The photometric redshift distribution of this sample is shown in Fig.~2.
The individual redshift probability distributions, $p(z)$,
of all these galaxies were used in the LF determination. In Tables A1 and A2 in the Appendix we provide 
a catalogue of the 100 most robust $z \simeq~6.5 - 12.0$ galaxies uncovered 
within the HUDF itself (i.e. from the UDF12 data), with positions, photometry, photometric redshifts (including 
uncertainties), total absolute magnitudes ($M_{1500}$) and cross-referencing to previous studies as appropriate. 
The corresponding information for the robust objects in the other seven survey fields analysed here is provided 
in the Tables A3-A9.

We note that the advance at faint magnitudes offered by the UDF12 dataset is clear, because 50 of the 100 
robust $z > 6.5$ HUDF galaxies tabulated in Table A1 have not been reported in any previous study.
Of these additional 50, colour-colour selection
(as described by Schenker et al. 2013) finds 23 sources. 
The remaining 27 galaxy candidates are only revealed by 
our photometric redshift analysis which exploits all of the 4-band WFC3/IR imaging. Note that, 
for simplicity, we have decided to report only $H_{160}$ magnitudes in Table A1, and so some {\it robust}
objects may appear to be surprisingly faint in $H_{160}$ because they are better detected 
in the shorter-wavelength WFC3/IR filters.

\subsection{Simulations}
The final result of the object selection process is a
catalogue of candidate high-redshift galaxies, each with an associated
photometric redshift probability density function. However, in order to accurately derived the galaxy luminosity function 
it is then necessary to employ detailed simulations to map between the derived and intrinsic properties of each candidate.

Following the methodology of McLure et al. (2009), we adopt a parametric 
model of the evolving high-redshift galaxy luminosity function in order 
to generate a realistic synthetic population of high-redshift galaxies. 
The model adopted here is a Schechter function with the three
parameters ($\phi^{\star}, M^{\star}_{UV}, \alpha$) evolving linearly
with redshift, changing from ($9.8\times 10^{-4}\rm{Mpc}^{-3},  -20.7, -1.65$) at
$z=4.5$ to ($2.3\times 10^{-4}\rm{Mpc}^{-3}, -20.1, -2.1$) at $z=9.0$. 
Although simple, this parameterization successfully reproduces the observed evolution
of the UV-selected galaxy luminosity function within the redshift interval $5<z<8$
(e.g. McLure et al. 2009, 2010; Bouwens et al. 2007, 2011; Bradley et al. 2012). 

For each survey field, the evolving luminosity function model is used to populate an input apparent magnitude - redshift 
grid ($m_{1500}-z$, where $m_{1500}$ is the apparent magnitude at 1500\AA), which is divided into cells of width 
$\delta z=0.05$ and $\delta m_{1500}=0.1$. For each simulated object, synthetic photometry is generated using 
Bruzual \& Charlot (2003) galaxy templates with a range of reddening designed to produce a population of objects whose 
distribution of UV slopes is centred on $\beta=-2$ with a small dispersion (c.f. Dunlop et al. 2012, 2013; Rogers, McLure \& Dunlop 2013). 
The synthetic galaxies were then injected into the real data, using the empirical measurement of the WFC3/IR and ACS PSF appropriate for each individual filter. It should be noted that simulations were also conducted in which the synthetic 
galaxies were modelled as spatially resolved, with half-light radii drawn from the distribution measured for $z=7-8$ LBGs by 
Oesch et al. (2010b). However, these simulations were not adopted because they were found to provide results virtually 
identical to the injection of PSFs and in the regime where the simulation results are most crucial for the luminosity function 
determination (i.e. $M_{1500}\geq -19$) several studies indicate that the UV-selected galaxy population is virtually unresolved at the
resolution of WFC3/IR (i.e. Oesch et al. 2010b; Grazian et al. 2012; Ono et al. 2013).

The simulated galaxy population was then selected and processed through our photometric redshift analysis in an identical fashion to the
real sample of high-redshift galaxy candidates. The resulting $p(z)$ distribution for each synthetic galaxy was then used to populate an 
output $m_{1500}-z$ grid. Under the assumption that the simulation provides a reasonable description of the actual 
high-redshift galaxy population, the ratio of the output and input $m_{1500}-z$ grids provides a mapping between the observed and 
intrinsic properties of the high-redshift galaxy population, which automatically accounts for selection efficiency, photometric redshift 
errors and flux boosting.

\begin{table}
\begin{center}
\caption{The results of our SWML determination of the $z=7$ and $z=8$ galaxy luminosity
functions. Columns 1 and 3 list the luminosity function bins adopted at $z=7$ and $z=8$ respectively (all bins are 0.5 magnitudes wide). Columns 2 and 4 list the individual values of $\phi_{k}$ and their corresponding uncertainties.}
\begin{tabular}{clcl}
\hline
 &\multicolumn{1}{c}{$z=7$}&&\multicolumn{1}{c}{$z=8$}\\
$M_{1500}$& $\phantom{00}\phi_{k}/$mag$^{-1}$Mpc$^{-3}$& $M_{1500}$& $\phantom{00}\phi_{k}/$mag$^{-1}$Mpc$^{-3}$\\
\hline
$-$21.00& \phantom{00}0.00003$\pm0.00001$    &     $-$21.25& \phantom{0}0.000008$\pm0.000003$ \\
$-$20.50& \phantom{00}0.00012$\pm0.00002$    &     $-$20.75& \phantom{00}0.00003$\pm0.000009$  \\    
$-$20.00& \phantom{00}0.00033$\pm0.00005$    &     $-$20.25& \phantom{000}0.0001$\pm0.00003$\\     
$-$19.50& \phantom{00}0.00075$\pm0.00009$    &     $-$19.75&\phantom{000}0.0003$\pm0.00006$\\  
$-$19.00& \phantom{000}0.0011$\pm0.0002$     &     $-$19.25&\phantom{000}0.0005$\pm0.00012$\\     
$-$18.50& \phantom{000}0.0021$\pm0.0006$     &     $-$18.75&\phantom{000}0.0012$\pm0.0004$ \\      
$-$18.00& \phantom{000}0.0042$\pm0.0009$     &     $-$18.25&\phantom{000}0.0018$\pm0.0006$ \\     
$-$17.50& \phantom{000}0.0079$\pm0.0019$     &     $-$17.75&\phantom{000}0.0028$\pm0.0008$\\  
$-$17.00& \phantom{0000}0.011$\pm0.0025$     &     $-$17.25&\phantom{000}0.0050$\pm0.0025$ \\
\hline\hline
\end{tabular}
\end{center}
\end{table}

\begin{figure*}
\includegraphics[width=16.0cm, angle=0]{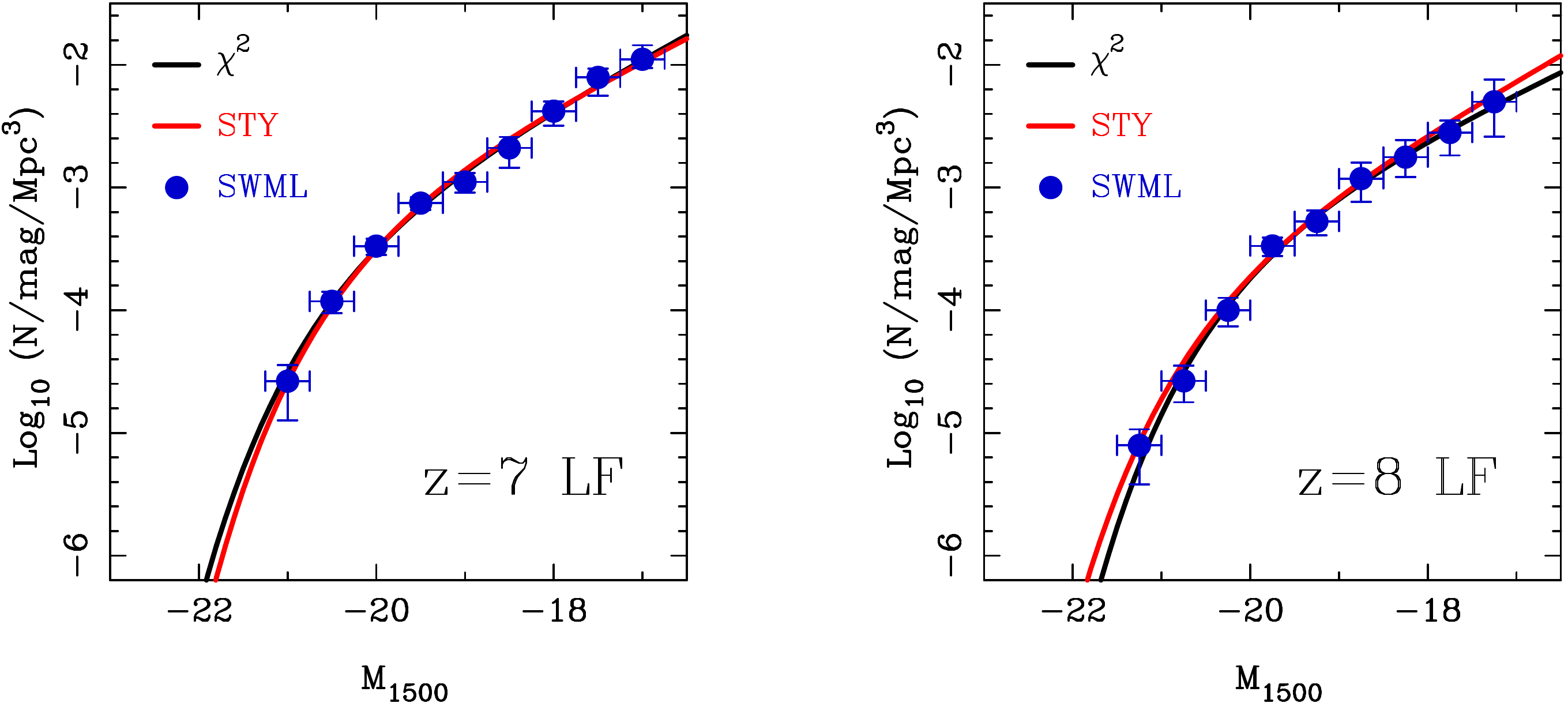}
\caption{The left-hand panel shows our new determination of the UV-selected galaxy luminosity function at redshift $z=7$. The data-points
have been derived using a combination of our photometric redshift analysis and an implementation of the 
step-wise maximum likelihood (SWML) method of Efstathiou, Ellis \& Peterson (1988); see text for details. The thick solid red 
line shows the best-fitting Schechter function derived using the parameteric STY maximum-likelihood technique applied simultaneously 
to the first seven survey fields listed in Table 1. For comparison, we also show a straight-forward $\chi^{2}-$fit to the binned SWML data (thick black line). The right-hand plot shows the same information for our new determination of the $z=8$ luminosity function. To derive the $z=8$ 
luminosity function we have also incorporated the information derived from our own reduction and analysis of the BoRG dataset.}
\end{figure*}

\section{Luminosity function estimation}
For each survey field, the $p(z)$ distributions for each high-redshift
galaxy candidate were used to populate an {\it observed} $m_{1500}-z$ plane, which was then corrected using the results of the
corresponding simulation to provide our best estimate of the
distribution of the observed high-redshift population on the {\it intrinsic} $m_{1500}-z$ plane. 
After aperture correcting the object magnitudes to total, it is this information that is then
used to estimate the galaxy luminosity function using two different
techniques.

The primary method is an implementation of  the non-parametric
step-wise maximum-likelihood (SWML) method of Efstathiou, Ellis \&
Peterson (1988). It is the results of this method which provide our
basic determination of $\phi(M_{1500},z)$, without relying on the
assumption that the luminosity function obeys a particular functional
form. However, in order to compare with previous results and to study
the redshift evolution of the galaxy luminosity function, it is also
desirable to derive parametric fits to the galaxy luminosity
function. Therefore, in order to derive Schechter-function fits to the galaxy luminosity function we have also
implemented a version of the parametric maximum-likelihood technique
of  Sandage, Tammann \& Yahil (1979); hereafter STY.

When defining the likelihood, both techniques rely on the assumption that, 
at a given redshift, the probability of observing a galaxy of a given luminosity can be defined as follows:
\begin{equation}
p_i \propto  \frac{\phi(L_i)}{\int_{\infty}^{L_{lim}} \phi(L) dL}
\end{equation}
\noindent
where $L_{lim}$ is the limiting luminosity of the survey. Ideally this should be implemented in the 
situation where each galaxy has a unique spectroscopic redshift and luminosity. In the absence of this information, 
our implementation does the next best thing and adopts the normalized probability density function for each 
high-redshift candidate, with the absolute UV magnitude ($M_{1500}$), calculated using a top-hat filter at 1500\AA\, in the rest-frame 
of the best-fitting SED template, re-calculated at each step within the $p(z)$. The overall best-fit is determined by maximizing the following likelihood:
\begin{equation}
\mathcal{L}=\prod_{j}\prod_{i} p_i
\end{equation}
\noindent
where the outer product symbol indicates that the maximum likelihood is calculated over $j$ separate survey fields, each with its own limiting luminosity.

One of the key strengths of both techniques is that they take no
account of the absolute number density of objects and should
therefore be relatively insensitive to the effects of
cosmic structure. However, as a result, it is necessary to determine  the
overall normalization of the resulting galaxy luminosity-function
estimates independently.  In each case, we have derived the luminosity function normalization by requiring that the
maximum-likelihood luminosity function estimates reproduce the
cumulative number counts of observed galaxies within the appropriate
redshift and absolute-magnitude intervals.

\begin{figure}
\includegraphics[width=8.0cm, angle=0]{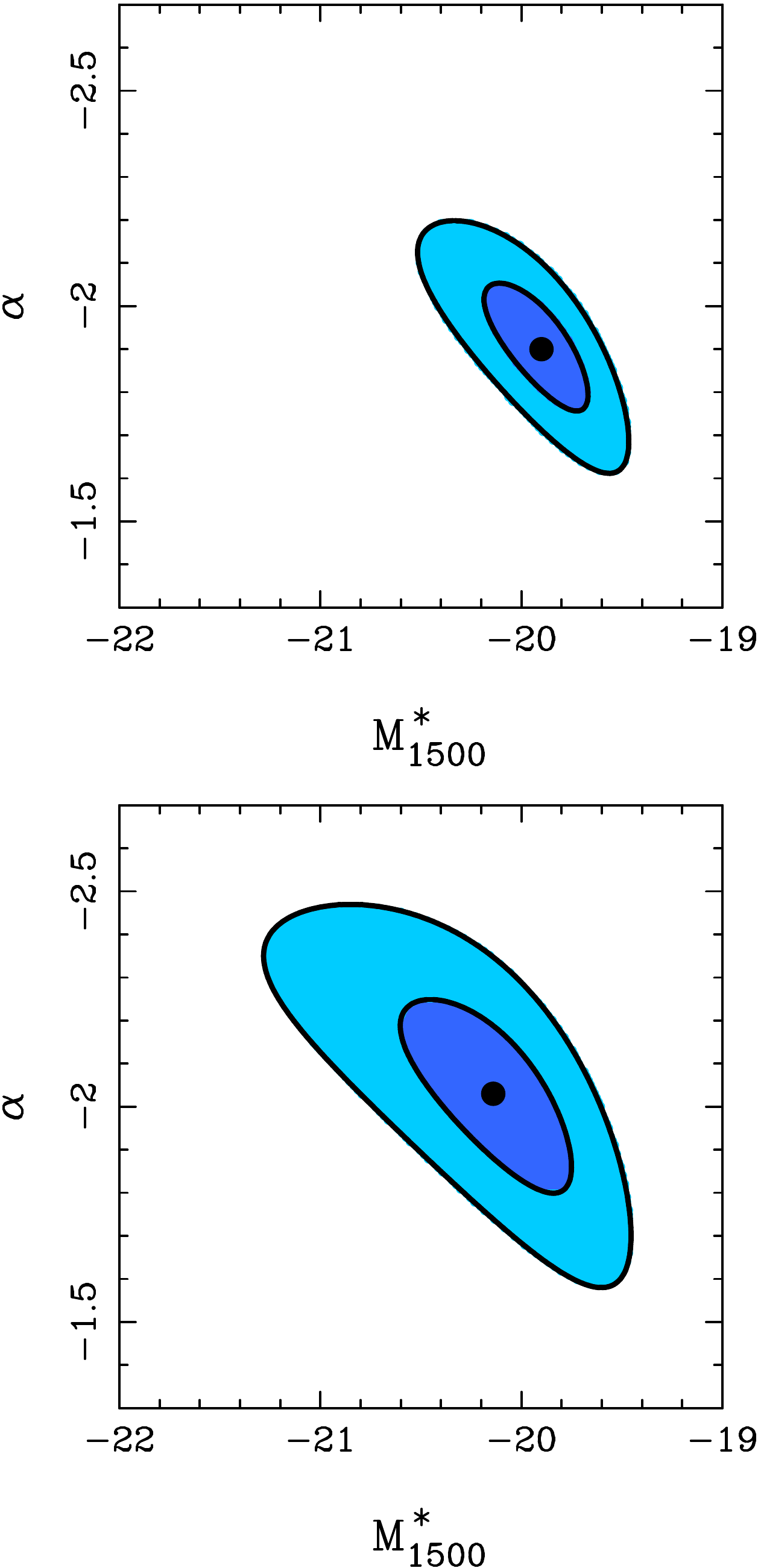}
\caption{The confidence intervals for the faint-end slope ($\alpha$) and characteristic absolute magnitude ($M^{\star}_{1500}$) derived 
from our STY maximum-likelihood fits to the galaxy luminosity function at $z=7$ (top) and $z=8$ (bottom). In each plot the $1\sigma$ and $2\sigma$ confidence intervals are illustrated 
by the dark and light-blue shaded areas respectively. It should be noted that the confidence intervals plotted are 
based on likelihood ratios and correspond to $\Delta \chi^{2}=1\,\&\, 4$ from the best overall fit (shown by the filled circle in each case). 
These specific confidence intervals have been chosen in order that the corresponding one parameter uncertainties can be calculated by 
projecting the contours onto the relevant axis.}
\end{figure}

\section{The galaxy luminosity function results}
In Fig. 3 we show our new determinations of the UV-selected galaxy luminosity functions at $z=7$ and $z=8$. In each 
panel the data-points have been derived using the SWML technique, and include error estimates which have been derived 
via boot-strap re-sampling of the underlying high-redshift galaxy sample. In Table 2 we provide the individual, step-by-step, 
SWML determinations of the $z=7$ and $z=8$ luminosity functions and their corresponding uncertainties. 

In each panel of Fig. 3 the thick red line 
shows our best-fitting Schechter function as derived via our implementation of the STY maximum-likelihood 
technique, which is constrained via a simultaneous fit to all relevant survey fields (as listed in Table 1). The confidence 
intervals on the derived faint-end slope and characteristic magnitude from our STY fits at $z=7$ and $z=8$ are shown in Fig. 4 and the 
best-fitting Schechter-function parameters are listed in Table 3.

As can be seen from the results presented in Fig. 3, the additional dynamic range in luminosity provided by combining the new 
UDF12 dataset with the wider area GOODS-S, CANDELS-UDS and BoRG datasets has allowed an accurate determination of the $z=7$ and 
$z=8$ luminosity functions spanning a factor $\geq 50$ in UV luminosity. In particular, the new UDF12 dataset has allowed us to constrain 
the form of the luminosity function as faint as $L \leq 0.1 L^{\star}$ for the first time. As can be seen from the results presented in 
Table 3, our new analysis has confirmed that the luminosity function remains extremely steep at $M_{1500}\geq -18$ at both $z=7$ and $z=8$, 
being consistent with $\alpha=-2.0$ in both cases. Moreover, our new analysis confirms and strengthens previous results which suggested 
that there is little evolution in $M^{\star}_{1500}$ between $z=7$ and $z=8$. In the next section we briefly compare our results with 
other relevant results in the literature, before proceeding to consider the evolution of the luminosity function from $z \simeq 6$ to $z \simeq 10$ and its implications for cosmic reionization.

\subsection{Comparison with previous results}
Although an exhaustive comparison with previous high-redshift luminosity function 
work in the literature is beyond the scope of this paper, it 
is instructive to compare the results derived here with those of other recent studies. 
Consequently, we will concentrate on a comparison between our new results and those of other recent {\it HST} studies which have attempted 
to fit all three Schechter-function parameters and their corresponding uncertainties. The details of the Schechter-function fits derived by 
the various different studies are provided in Table 3.

Given that it also includes the new UDF12 data, it is obviously of interest to compare our results to those of the companion drop-out analysis of the 
$z=7$ and $z=8$ luminosity functions performed by our team (Schenker et al. 2013). It can be seen from Table 3 that at $z=7$ our results are in
good agreement, with the extra luminosity leverage provided by UDF12 leading both studies to conclude that the faint-end slope is $\alpha=-1.9$, 
with a small uncertainty. The overall uncertainties on the Schechter-function parameters derived in this work are slightly tighter than in 
Schenker et al. (2013), which is expected given that we have analysed a larger survey area. At $z=8$ our results are again consistent with those 
published in Schenker et al. (2013), particularly in term of the faint-end slope, although the value of 
$M_{1500}^{\star}$ derived by Schenker et al. is somewhat brighter than found here.

Before the start of the UDF12 imaging campaign, the most comprehensive study of the $z=7$ and $z=8$ luminosity functions was 
performed by Bouwens et al. (2011a), who combined a dropout analysis of the WFC3/IR imaging in the HUDF09 and ERS (total area 53 arcmin$^2$) 
with constraints provided by various wider-area datasets. It can be seen from Table 3 that, in terms of derived luminosity function parameters, 
there is actually very good agreement between the new results derived here and those of Bouwens et al. (2011a). At some level this may be 
slightly fortuitous, given that the agreement between the Schechter function parameters is better than that between our respective binned SWML 
results and, as noted by Bouwens et al. (2011a), the small area of their study meant that their $z=8$ faint-end slope could have been as 
shallow as $\alpha=-1.67\pm{0.40}$ depending on the inclusion/exclusion of two bright candidates in HUDF09-2. However, irrespective of this, it 
is clear that the fundamental advantage of our new analysis is the ability to better constrain the faint-end slope at both redshifts, using 
the deeper UDF12 imaging.

Much of the recent work in the literature has been focused on trying to improve our knowledge of the $z=8$ luminosity function. Within this 
context, two recent studies by Oesch et al. (2012b) and Bradley et al. (2012) have investigated the form of the $z=8$ luminosity function by 
combining the faint-end results of Bouwens et al. (2011a), with improved constraints at the bright end. In the case of Bradley et al. (2012) the 
bright-end information is provided by their drop-out analysis of the BoRG WFC3/IR parallel observations, whereas in Oesch et al. (2012b) the 
bright-end constraints are provided by a drop-out analysis of the CANDELS DEEP+WIDE imaging in GOODS-S (total area 95 arcmin$^2$). It can 
be seen from the results presented in Table 3 that, in terms of derived luminosity-function parameters, there is very good agreement 
between the new results derived here and those of Oesch et al. (2012b) and Bradley et al. (2012), with all studies seemingly converging on a 
steep faint-end slope of $\alpha\simeq -2.0$ and $M^{\star}_{1500}\simeq -20.1$.

Overall, the comparison between derived luminosity function parameters shown in Table 3 is therefore 
highly encouraging, especially given the different datasets, 
reductions and analysis techniques adopted by the various different studies. However, due to the fact that the current study (together with 
Schenker et al. 2013) exploits the deeper imaging provided by UDF12 and, uniquely, incorporates WFC3/IR imaging covering a wider area than all previous 
studies (including GOODS-S, CANDELS-UDS and BoRG) we are confident that 
the luminosity-function determination provided here is 
the most accurate currently available at these redshifts.

\begin{table}
\caption{The Schechter-function parameters for the $z=7$ and $z=8$ galaxy
luminosity function derived by various recent {\it HST} studies. The first column lists the name of
the study and columns two to four list the Schechter-function parameters and their quoted uncertainties. 
The units of $\phi^{\star}$ are Mpc$^{-3}$.}
\begin{tabular}{lccc}
\hline
Study & $M_{1500}^{\star}$ & $\log(\phi^{\star})$  & $\alpha$\\
\hline
\multicolumn{1}{l}{$z=7$}&&&\\[2ex]
This work  & $-19.90^{+0.23}_{-0.28}$ & $-2.96^{+0.18}_{-0.23}$ & $-1.90^{+0.14}_{-0.15}$\\[2ex]
Schenker et al. (2013)  & $-20.14^{+0.36}_{-0.48}$ & $-3.19^{+0.27}_{-0.24}$ & $-1.87^{+0.18}_{-0.17}$\\[2ex]
Bouwens et al. (2011a)  &$-20.14^{+0.26}_{-0.26}$ & $-3.07^{+0.26}_{-0.26}$ & $-2.01^{+0.21}_{-0.21}$\\[2ex]
\hline
\multicolumn{1}{l}{$z=8$}&&&\\[2ex]
This work  & $-20.12^{+0.37}_{-0.48}$ & $-3.35^{+0.28}_{-0.47}$ & $-2.02^{+0.22}_{-0.23}$\\[2ex]
Schenker et al. (2013) &$-20.44^{+0.47}_{-0.35}$ & $-3.50^{+0.35}_{-0.32}$ & $-1.94^{+0.21}_{-0.24}$\\[2ex]
Bouwens et al. (2011a)  &$-20.10^{+0.52}_{-0.52}$ & $-3.23^{+0.43}_{-0.43}$ & $-1.91^{+0.32}_{-0.32}$\\[2ex]
Oesch et al. (2012b)    &$-20.04^{+0.44}_{-0.48}$ & $-3.30^{+0.38}_{-0.46}$ & $-2.06^{+0.35}_{-0.28}$\\[2ex]
Bradley et al. (2012)  &$-20.26^{+0.29}_{-0.34}$ & $-3.37^{+0.26}_{-0.29}$ & $-1.98^{+0.23}_{-0.22}$\\[2ex]
\hline\hline
\end{tabular}
\end{table}

\begin{figure}
\includegraphics[width=8.0cm, angle=0]{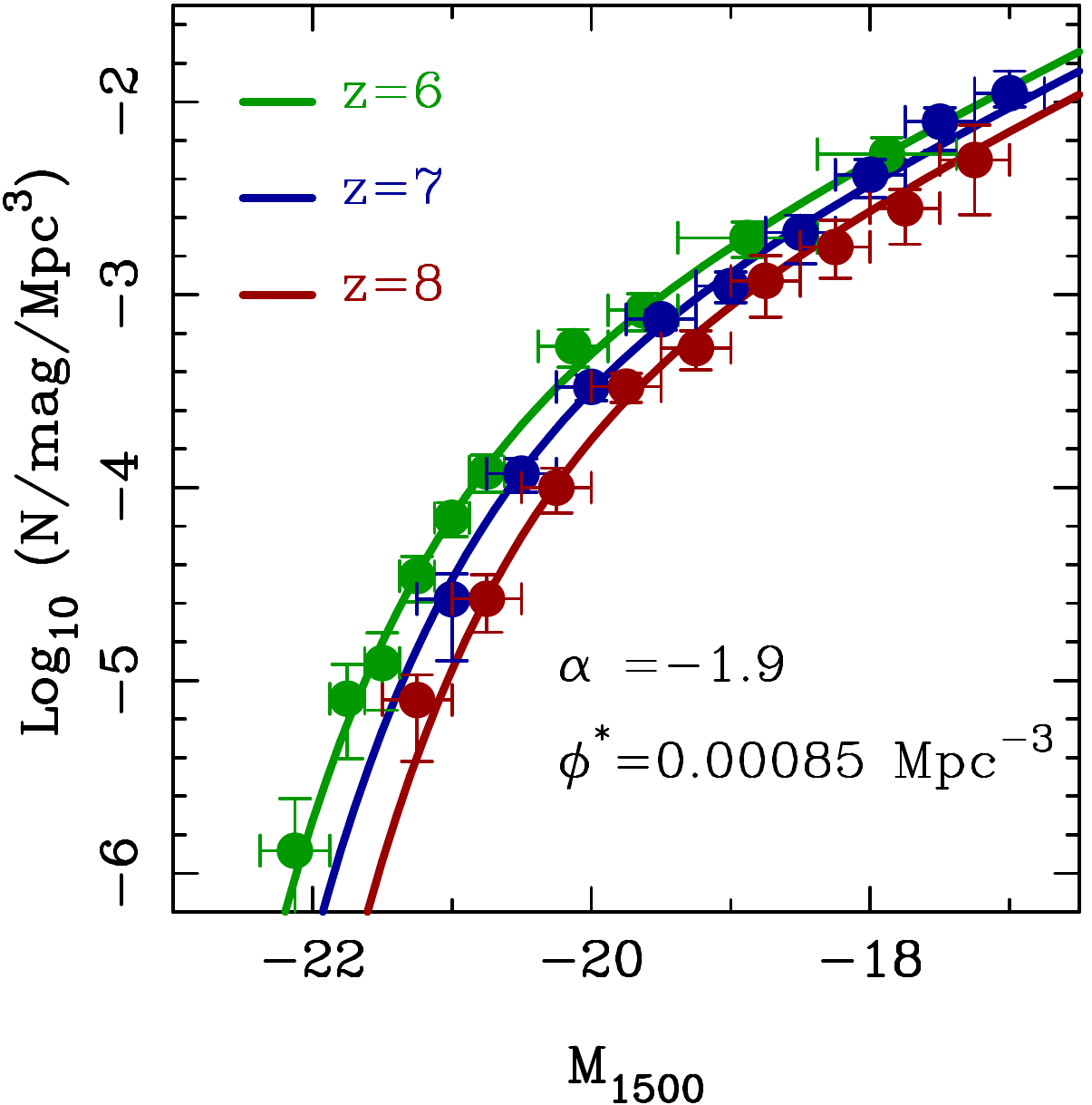}
\caption{An illustration that pure luminosity evolution can provide an acceptable fit to the observed evolution of the galaxy
luminosity function over the redshift interval $6<z<8$. The blue and red data-points show the SWML determination of the galaxy luminosity 
function at $z=7$ and $z=8$ from this work, while the green data-points show the determination of the $z=6$ luminosity function from 
McLure et al. (2009). The corresponding curves show the results of fitting the binned LF data with Schechter functions where the 
faint-end slope and overall normalization have been held constant at representative values ($\alpha=-1.9$ and $\phi^{\star}=0.00085$ Mpc$^{-3}$ 
respectively) but the characteristic magnitude ($M_{1500}^{\star}$) has been allowed to float. In this scenario $M_{1500}^{\star}$ evolves by $\simeq 0.3$ magnitudes per $\Delta z=1$ interval, changing from $M_{1500}^{\star}\simeq-20.3$ at $z=6$ to $M_{1500}^{\star}\simeq-19.7$ at $z=8$. It can be seen that this simple parameterization is capable of satisfactorily reproducing the observed data.}
\end{figure}

\subsection{The evolution of the luminosity function}
Several previous studies have concluded that the evolution of the galaxy luminosity function over the redshift range $5<z<7$ can be well 
described as pure luminosity evolution (e.g. Bouwens et al. 2007; McLure et al. 2009; Bouwens et al. 2011a). It is clearly of some interest to 
investigate whether or not the evolution of the luminosity function from $z=7-8$ remains consistent with this apparently simple picture. 

Some insight into this question can be gained by examining the confidence intervals on the faint-end slope and characteristic magnitude at $z=7$ 
and $z=8$ shown in Fig. 4. It can immediately be seen from Fig. 4 that our new analysis provides little evidence for a significant change 
in $M^{\star}_{1500}$ or $\alpha$ over the redshift inteval $z=7-8$ and in fact, the best-fitting Schechter-function parameters (see Table 3)
suggest that the dominate change is a factor of $\simeq 2.5$ drop in $\phi^{\star}$ between $z=7$ and $z=8$. We note here that 
Bouwens et al. (2011a) also commented that some of the $z=7-8$ evolution may be explained by a change in $\phi^{\star}$, but concluded that 
the uncertainties were too large to be confident. Although our improved determinations of the $z=7$ and $z=8$ luminosity functions 
strengthen the suggestion that $\phi^{\star}$ is changing within the redshift range $7<z<8$, the results presented in Fig. 5 indicate that 
the available data are still insufficient to rule out pure luminosity evolution.

Given that pure luminosity evolution provides such a good description of the luminosity function evolution over the redshift 
interval $4<z<7$ (Bouwens et al 2011a), in Fig. 5 we explore whether a simple luminosity evolution parameterization can continue to 
provide an adequate description of the observed evolution at $z\geq 6$. To investigate this issue we simply fit 
Schechter functions to the binned SWML luminosity-function data at $z=6, 7\,\&\,8$, allowing $M^{\star}_{1500}$ to float as a free parameter, 
but keeping $\phi^{\star}$ and $\alpha$ fixed at representative values ($\alpha=-1.9$ and $\phi^{\star}=0.00085$ Mpc$^{-3}$ respectively). 
In this simplied scenario we find that $M_{1500}^{\star}$ evolves by $\simeq 0.3$ magnitudes per $\Delta z=1$ interval, changing from 
$M_{1500}^{\star}\simeq-20.3$ at $z=6$ to $M_{1500}^{\star}\simeq-19.7$ at $z=8$. It is immediately clear from Fig. 5 that, within the 
constraints of the current data, it is still perfectly possible to reproduce the observed luminosity function data in the redshift 
interval $6<z<8$ with pure luminosity evolution alone. 

\subsection{The galaxy luminosity function at $\bmath{z=9}$}
In addition to the quadrupling of the available $Y_{105}$ imaging in the HUDF, the key advantage provided by the new UDF12 dataset is 
the addition of ultra-deep imaging in the previously unexploited $J_{140}$ filter. The availability of the 
new $J_{140}$ imaging provides the first real opportunity to constrain the faint end of the $z\simeq 9$ luminosity function, 
simply because, in the redshift interval $8.5<z<9.5$, the $J_{140}$ and $H_{160}$ imaging still provide two filters long-ward of the redshifted Lyman break.
Although there is significant overlap ($\simeq 2/3$) between the $J_{140}$ and $H_{160}$ filters, the availability of two images with different 
noise properties is invaluable for ruling out spurious sources and avoiding the notorious problems associated with single-band detections long-ward 
of the Lyman break.

In Fig. 6 we show our SWML determination of the $z=9$ galaxy luminosity function, which is derived entirely from the data available 
in the HUDF itself. For comparison, we also show in Fig. 6 our SWML determination of the $z=8$ luminosity function, which is identical 
that shown in Fig. 3. Although it is clearly not sensible to draw strong conclusions from two low signal-to-noise luminosity-function bins, 
the $z=9$ data-points shown in Fig. 6 (and listed in Table 4) immediately suggest that there is no dramatic fall in the volume density 
of $M_{1500} \simeq -18$ galaxies between $z=8$ and $z=9$. The solid and dashed lines in Fig. 6 show the result of Schechter function fits to the two $z=9$ data-points 
under the assumption that the evolution from $z=8$ to $z=9$ is either purely luminosity evolution (solid line) or purely density 
evolution (dashed line). In both cases it is assumed that the luminosity function faint-end slope remains unchanged at $\alpha=-2.02$.

It can clearly be seen from Fig. 6 that it is impossible to say anything meaningful about the {\it form} of any evolution of the 
galaxy luminosity function between $z=8$ and $z=9$. However, the two alternative Schechter-function fits do demonstrate that, 
provided the faint-end slope does not change substantially between $z=8$ and $z=9$, irrespective of the form of the evolution 
the resulting integrated UV luminosity density is likely to be very similar.

\begin{table}
\begin{center}
\caption{The results of our SWML determination of two luminosity bins on the $z=9$ galaxy luminosity
function. Column 1 lists the adopted luminosity function bins (both are 0.5 magnitudes wide) and column 2 lists the individual values 
of $\phi_{k}$ and their corresponding uncertainties.}
\begin{tabular}{cl}
\hline
$M_{1500}$& $\phi_{k}/$mag$^{-1}$Mpc$^{-3}$\\
\hline
$-$18.00& \phantom{00}0.0016$\pm0.0007$\\  
$-$17.50& \phantom{00}0.0021$\pm0.0009$\\
\hline\hline
\end{tabular}
\end{center}
\end{table}

\begin{figure}
\includegraphics[width=8.0cm, angle=0]{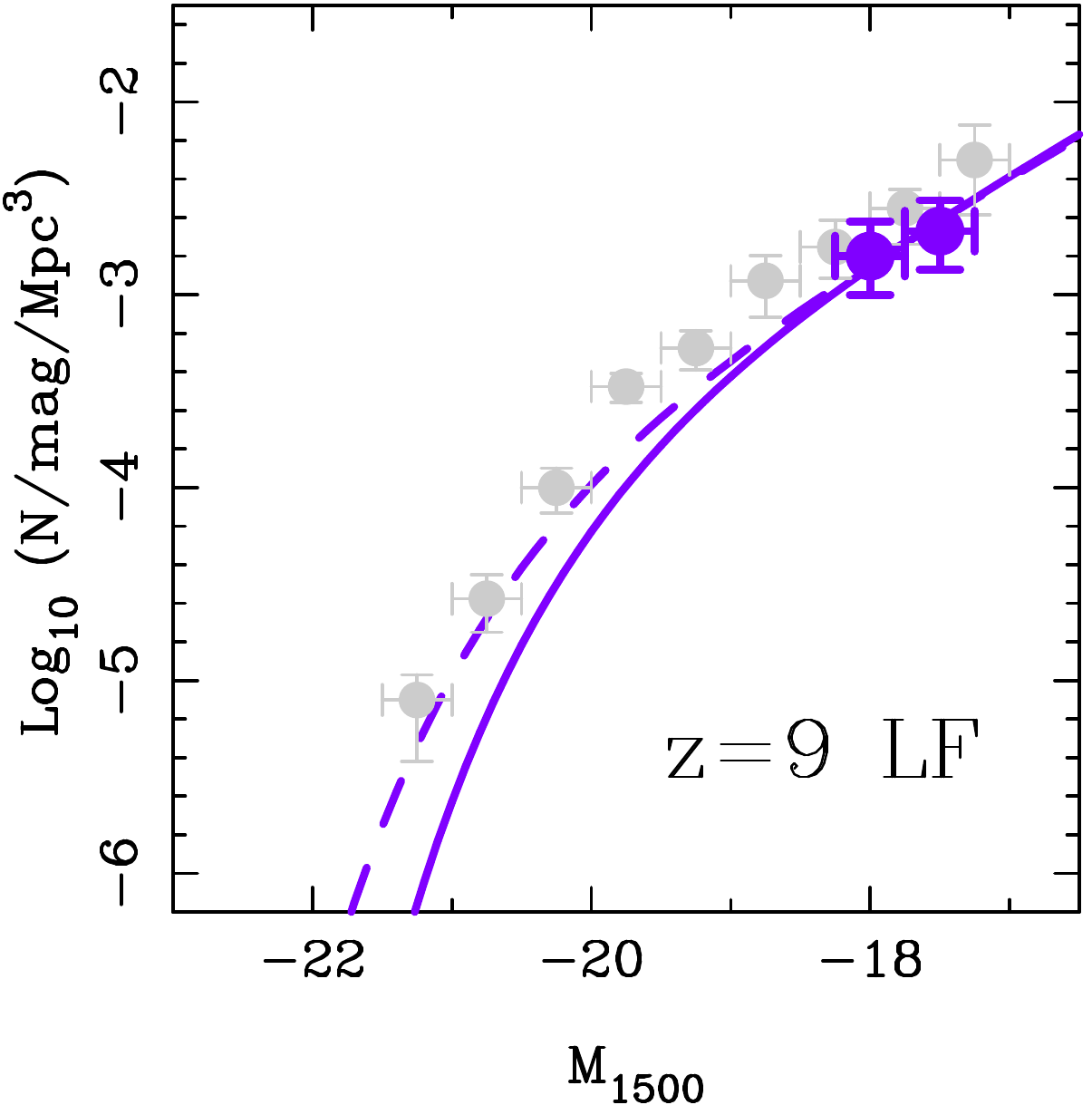}
\caption{The large purple data-points show our SWML estimate of the $z \simeq 9$ luminosity function derived from the UDF12 dataset. 
For context, in grey we also show our estimate of the $z=8$ luminosity function. The solid and dashed lines show fits to the $z=9$ 
data-points under the assumption that the luminosity function evolves from $z=8$ to $z=9$ via pure luminosity (solid) or pure density 
(dashed) evolution respectively (both assume the faint-end slope remains fixed at the $z=8$ value). 
Although it is currently impossible to differentiate between them, it is clear that both scenarios would lead to very 
similar integrated UV luminosity densities (see text for a discussion).}
\end{figure}

\begin{figure}
\includegraphics[width=8.0cm, height=15.0cm, angle=0]{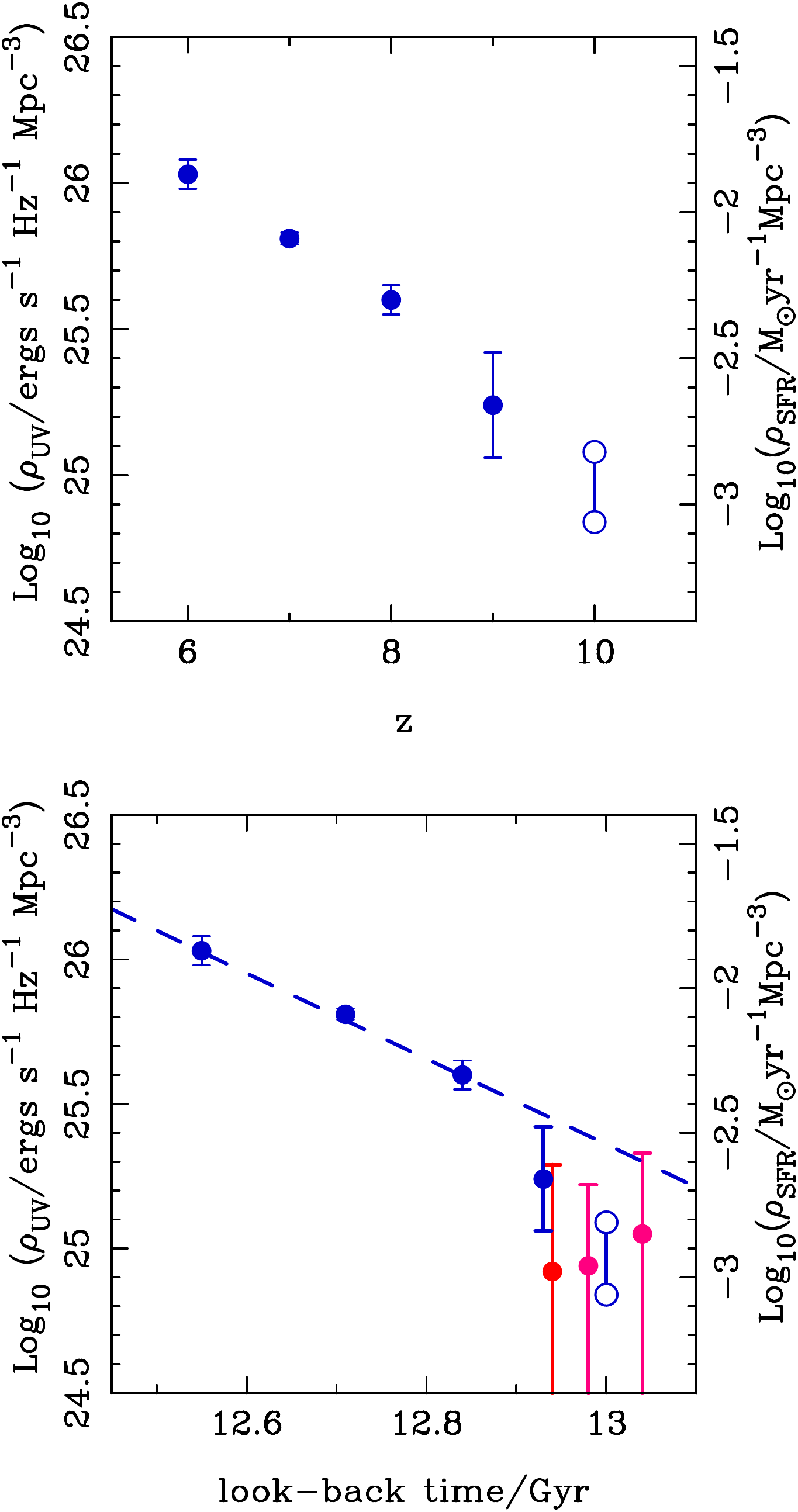}
\caption{The evolution of the observed UV luminosity density as a function of redshift (top) and look-back time (bottom).
The data-points at $6<z<9$ show the results of integrating the galaxy luminosity function to an absolute magnitude limit of $M_{1500}=-17.7$, where the
luminosity function parameters are taken from McLure et al. (2009) at $z=6$ and from this work for $z=7-9$. 
The upper and lower data-points at $z=10$ are derived by assuming that the evolution of the luminosity function from $z=8-9$ continues as either 
pure luminosity (lower) or pure density (upper) evolution respectively. The dashed line shown in the bottom panel is a linear fit to the evolution of the UV 
luminosity density between $z=6$ and $z=8$. In the bottom panel the two pink data-points are the recent UV luminosity estimates from Coe et al. (2013), while 
the red data-point is the estimate from Bouwens et al. (2013). In each panel the right-hand axis shows how UV luminosity density converts to star-formation 
density assuming a Salpeter IMF and the UV-to-SFR conversion of Madua, Pozzetti \& Dickinson (1998).}
\end{figure}

\section{Evolving luminosity density and cosmic star-formation history}

Given our improved knowledge of the evolving galaxy luminosity function, it is clearly of interest to briefly 
explore the implications for the evolution of the observed UV luminosity density, which has a direct impact on determining 
whether the observed high-redshift galaxy population can reionize the Universe. For a thorough review of the constraints which 
could be placed on reionization pre-UDF12, the reader is referred to Robertson et al. (2010) and Finkelstein et al. (2012).

In Fig. 7 we provide a new calculation of the evolution of the observed UV luminosity density based on the McLure et al. (2009) 
determination of the $z=6$ luminosity function and the new luminosity function determinations derived in this work at $z=7-9$.
In all cases the datapoints show the results of integrating the appropriate luminosity function down to an absolute magnitude 
of $M_{1500}=-17.7$, in order to allow straightforward comparison with previous work. In both panels the right-hand axis shows how the UV luminosity density converts into a star-formation rate density under the assumption of a Salpeter IMF and the conversion between UV luminosity and star-formation rate prescribed by Madau, Pozzetti \& Dickinson (1998). In the top panel we show the evolution in UV luminosity density as a function of redshift, whereas in the bottom panel we show the evolution as a function of look-back time. The upper and lower datapoints at $z=10$ show how the UV luminosity density changes if the luminosity function evolution from $z=8-9$ continues 
to $z=10$ as either pure luminosity (lower) or pure density evolution (upper) respectively. No dust corrections have been applied to any of the datapoints.

It can be seen from Fig. 7 that our data indicate that there is an order of magnitude increase in the observed UV luminosity density 
over the 450 Myr period between $z=10$ and $z=6$. Moreover, from the bottom panel of Fig. 7 it can be seen that the increase 
in UV density is very close to linear with cosmic time between $z=8$ and $z=6$. However, our new information at $z\simeq 9$ from the  UDF12 dataset provides some evidence that the fall-off in UV density at $z\geq 8$ is steeper than a linear trend 
with cosmic time, particularly if the galaxy luminosity function continues to evolve primary via luminosity evolution. We note 
that in this regard our results are consistent with Oesch et al. (2012a) and the recent results based on the CLASH clusters campaign 
published by Coe et al. (2013) and Bouwens et al. (2013), which are plotted in the bottom panel of Fig. 7. A full discussion of the implications of these results within the context of cosmic reionization can be found in Robertson et al. (2013).

\section{Conclusions}
By combining the extreme near-IR depth provided by the UDF12 campaign with extensive wider area WFC3/IR imaging data, it has been 
possible to study the high-redshift galaxy luminosity function using galaxy 
samples selected from a total area of $\simeq 480$\,arcmin$^2$, which span a factor of $\geq 50$ in luminosity and a factor of 
$\simeq 1000$ in number density. Based on this unique dataset it has been possible to determine the most accurate measurement 
to date of the $z=7$ and $z=8$ galaxy luminosity functions and the first meaningful constraints on the galaxy luminosity 
function at $z\simeq 9$. The principal results and conclusions of our study can be summarised as follows:

\begin{enumerate}

\item{The extra depth provided by the UDF12 dataset has allowed us to demonstrate that the faint-end slope of the 
galaxy luminosity function at $z=7$ and $z=8$ remains extremely steep, down to $M_{1500}=-16.75$ and $M_{1500}=-17.00$ 
respectively. Based on fitting Schechter functions, our 
formal constraints on the faint-end slope are $\alpha = -1.90^{+0.14}_{-0.15}$ at $z=7$ and $\alpha = -2.02^{+0.22}_{-0.23}$ at $z=8$.}

\item{The results of our Schechter function fits strengthen previous suggestions that the form of the evolution of the 
luminosity function between $z=7$ and $z=8$ is more akin to density evolution, rather than the apparent luminosity evolution observed at redshifts $z=5-7$.}

\item{However, even with the extra leverage provided by the UDF12 dataset, we conclude that it is not possible to differentiate 
between luminosity and density evolution between $z=7$ and $z=8$. In fact, we demonstrate that it is perfectly possible to provide an
adequate description of the observed luminosity function data between $z=6$ and $z=8$ under the assumption of pure luminosity evolution alone.}

\item{The unique nature of the UDF12 dataset has allowed us to place the first meaningful constraints on the faint end of the 
galaxy luminosity function at $z=9$. Taken at face value, these initial results suggest that, at least at $M_{1500}\simeq -18$, there is not 
a dramatic fall-off in the volume density of faint galaxies between $z=8$ and $z=9$.}

\item{Based on our determinations of the galaxy luminosity function within the redshift interval $6<z<9$, we briefly explore 
the evolution of the observed UV luminosity density. Our results indicate that there is an order of magnitude increase in the 
UV luminosity density over the redshift range $6<z<10$ and that between $z=8$ and $z=6$ the UV luminosity density increases 
linearly with cosmic time. However, our new results at $z\simeq 9$, together with recent results from the literature, suggest that the fall-off in UV luminosity density at $z\geq 8$ is steeper than would be expected for a linear trend with cosmic time.}

\end{enumerate}

\section*{acknowledgements}
RJM acknowledges the support of the European Research Council via the award of a 
Consolidator Grant, and the support of the Leverhulme Trust via the award of a Philip Leverhulme research prize.  
JSD, RAAB, TAT, and VW acknowledge the support of the European Research Council via the award of an
Advanced Grant to JSD. JSD also acknowledges the support of the Royal Society via a Wolfson Research Merit award.
ABR and EFCL acknowledge the support of the UK Science \& Technology Facilities Council.
US authors acknowledge financial support from the Space Telescope Science Institute under 
award HST-GO-12498.01-A. 
SRF is partially supported by the David and Lucile Packard Foundation.
SC acknowledges the support of the European Commission through the Marie Curie Initial Training Network ELIXIR. 
This work is based in part on observations made with the NASA/ESA {\it Hubble Space Telescope}, which is operated by the Association 
of Universities for Research in Astronomy, Inc, under NASA contract NAS5-26555.
This work is also based in part on observations made with the {\it Spitzer Space Telescope}, which is operated by the Jet Propulsion Laboratory, 
California Institute of Technology under NASA contract 1407. 

{}

\newpage
\begin{appendix}
\section{Robust high-redshift candidates}
In Table A1 we provide the coordinates, photometric redshifts and absolute UV magnitudes ($M_{1500}$) for our N=100 robust
$z\geq 6.5$ galaxy candidates in the HUDF field. In Table A2 we provide the $z_{850}, Y_{105}, J_{125}, J_{140}$ and $H_{160}$ photometry for the same sample. In Tables A3-A9 we provide the coordinates, photometry, photometric redshifts and absolute UV magnitudes ($M_{1500}$) for our robust
$z\geq 6.5$ galaxy candidates in the HUDF09-1, HUDF09-2, ERS, CANDELS GS-DEEP, CANDELS GS-WIDE, CANDELS UDS and BoRG fields. In all cases we quote a minimum photometric error of $\pm0.1$ magnitudes, even for those objects which are detected at $\geq 10\sigma$ significance.

\begin{table*}
\caption{Candidate $z\geq 6.5$ galaxies in the HUDF. Column one lists the candidate names and columns two and three list the coordinates. Columns four and five list 
the best-fitting photometric redshift and the corresponding $1\sigma$ uncertainty. Column six lists the total absolute UV magnitude, measured using a top-hat 
filter at 1500\AA\ in the rest-frame of the best-fitting galaxy SED template. Column seven lists the total apparent $H_{160}$ magnitude (corrected to total assuming a point source). Column eight
gives references to previous discoveries of objects: (1)~McLure et al. (2010), (2)~Oesch et al. (2010a), (3)~Finkelstein et al. (2010),
(4)~Bunker et al. (2010), (5)~Yan et al. (2010), (6)~Bouwens et al. (2010), (7)~Wilkins et al. (2011) (8)~Lorenzoni et al. (2011), (9)~Bouwens et al. (2011a), (10)~Bouwens et al. (2011a) potential, 
(11)~McLure et al. (2011), (12)~Bouwens et al. (2011b) (13)~Schenker et al. (2013).}
\begin{tabular}{lccccccc}
\hline
Name & RA(J2000) & Dec(J2000) & $z_{phot}$ &$\Delta z$ & $M_{1500}$ & $H_{160}$ & References\\
\hline
  UDF12-3999-6197 & 03:32:39.99 & $-$27:46:19.7 & 6.5 & 6.2$-$7.1 & $-$17.5 & 28.8$^{+0.1}_{-0.1}$&                  \\[1ex]
  UDF12-3696-5536 & 03:32:36.96 & $-$27:45:53.6 & 6.5 & 6.1$-$6.9 & $-$17.5 & 29.5$^{+0.3}_{-0.2}$&              9,13\\[1ex]
  UDF12-3677-7536 & 03:32:36.77 & $-$27:47:53.6 & 6.5 & 6.4$-$6.6 & $-$19.0 & 27.8$^{+0.1}_{-0.1}$&   1,2,3,7,9,11,13\\[1ex]
  UDF12-3897-8116 & 03:32:38.97 & $-$27:48:11.6 & 6.5 & 6.1$-$7.0 & $-$17.4 & 29.4$^{+0.3}_{-0.2}$&                13\\[1ex]
  UDF12-4120-6561 & 03:32:41.20 & $-$27:46:56.1 & 6.5 & 6.3$-$6.9 & $-$16.9 & $>30.2$             &                  \\[1ex]
  UDF12-3515-7257 & 03:32:35.15 & $-$27:47:25.7 & 6.5 & 6.3$-$6.8 & $-$17.4 & 29.7$^{+0.4}_{-0.3}$&                  \\[1ex]
  UDF12-3909-6092 & 03:32:39.09 & $-$27:46:09.2 & 6.5 & 6.3$-$6.8 & $-$17.5 & 29.5$^{+0.3}_{-0.2}$&                13\\[1ex]
  UDF12-3865-6041 & 03:32:38.65 & $-$27:46:04.1 & 6.6 & 6.3$-$6.8 & $-$17.8 & 29.4$^{+0.3}_{-0.2}$&                  \\[1ex]
  UDF12-3702-5534 & 03:32:37.02 & $-$27:45:53.4 & 6.6 & 6.2$-$6.9 & $-$17.2 & 30.2$^{+0.6}_{-0.4}$&                 9\\[1ex]
  UDF12-3922-6148 & 03:32:39.22 & $-$27:46:14.8 & 6.6 & 6.2$-$7.0 & $-$17.2 & 29.7$^{+0.4}_{-0.3}$&                13\\[1ex]
  UDF12-3736-6245 & 03:32:37.36 & $-$27:46:24.5 & 6.6 & 6.3$-$6.9 & $-$17.7 & 29.2$^{+0.2}_{-0.2}$&              9,13\\[1ex]
  UDF12-4379-6511 & 03:32:43.79 & $-$27:46:51.1 & 6.6 & 6.4$-$6.8 & $-$17.7 & 29.4$^{+0.3}_{-0.2}$&                13\\[1ex]
  UDF12-3859-6521 & 03:32:38.59 & $-$27:46:52.1 & 6.6 & 6.4$-$6.8 & $-$17.8 & 29.2$^{+0.2}_{-0.2}$&              9,13\\[1ex]
  UDF12-4202-7074 & 03:32:42.02 & $-$27:47:07.4 & 6.6 & 6.3$-$7.0 & $-$17.6 & 29.1$^{+0.2}_{-0.2}$&                13\\[1ex]
  UDF12-3638-7163 & 03:32:36.38 & $-$27:47:16.3 & 6.6 & 6.4$-$6.7 & $-$18.7 & 28.2$^{+0.1}_{-0.1}$& 1,2,3,4,5,7,9,11,13\\[1ex]
  UDF12-4254-6481 & 03:32:42.54 & $-$27:46:48.1 & 6.6 & 6.3$-$6.9 & $-$17.1 & 30.2$^{+0.7}_{-0.4}$&                  \\[1ex]
  UDF12-4058-5570 & 03:32:40.58 & $-$27:45:57.0 & 6.6 & 6.3$-$6.8 & $-$18.0 & 29.1$^{+0.2}_{-0.2}$&                  \\[1ex]
  UDF12-3858-6150 & 03:32:38.58 & $-$27:46:15.0 & 6.6 & 5.9$-$7.4 & $-$17.1 & 29.7$^{+0.4}_{-0.3}$&                  \\[1ex]
  UDF12-4186-6322 & 03:32:41.86 & $-$27:46:32.2 & 6.6 & 6.2$-$7.0 & $-$17.8 & 29.0$^{+0.2}_{-0.2}$&                 9\\[1ex]
  UDF12-4144-7041 & 03:32:41.44 & $-$27:47:04.1 & 6.6 & 6.1$-$7.0 & $-$17.4 & 29.6$^{+0.4}_{-0.3}$&                13\\[1ex]
  UDF12-4288-6261 & 03:32:42.88 & $-$27:46:26.1 & 6.6 & 6.3$-$6.9 & $-$17.4 & 30.1$^{+0.6}_{-0.4}$&                13\\[1ex]
  UDF12-3900-6482 & 03:32:39.00 & $-$27:46:48.2 & 6.6 & 6.4$-$7.0 & $-$18.3 & 28.2$^{+0.1}_{-0.1}$&                  \\[1ex]
  UDF12-4182-6112 & 03:32:41.82 & $-$27:46:11.2 & 6.7 & 6.4$-$7.1 & $-$18.1 & 28.5$^{+0.1}_{-0.1}$&     3,7,7,9,11,13\\[1ex]
  UDF12-4268-7073 & 03:32:42.68 & $-$27:47:07.3 & 6.7 & 6.4$-$7.0 & $-$18.3 & 28.5$^{+0.1}_{-0.1}$&                13\\[1ex]
  UDF12-3734-7192 & 03:32:37.34 & $-$27:47:19.2 & 6.7 & 6.4$-$6.9 & $-$18.0 & 29.1$^{+0.2}_{-0.2}$&                13\\[1ex]
  UDF12-3968-6066 & 03:32:39.68 & $-$27:46:06.6 & 6.7 & 6.2$-$7.2 & $-$17.1 & $>30.2$             &                 9\\[1ex]
  UDF12-4219-6278 & 03:32:42.19 & $-$27:46:27.8 & 6.7 & 6.6$-$6.9 & $-$19.2 & 27.7$^{+0.1}_{-0.1}$&     1,3,7,9,11,13\\[1ex]
  UDF12-3796-6020 & 03:32:37.96 & $-$27:46:02.0 & 6.7 & 6.3$-$7.1 & $-$17.5 & 29.5$^{+0.3}_{-0.2}$&                  \\[1ex]
  UDF12-3675-6447 & 03:32:36.75 & $-$27:46:44.7 & 6.7 & 6.4$-$7.3 & $-$18.1 & 28.7$^{+0.1}_{-0.1}$&                  \\[1ex]
  UDF12-3744-6513 & 03:32:37.44 & $-$27:46:51.3 & 6.7 & 6.6$-$6.9 & $-$19.0 & 28.1$^{+0.1}_{-0.1}$& 1,2,3,5,7,9,11,13\\[1ex]
  UDF12-4160-7045 & 03:32:41.60 & $-$27:47:04.5 & 6.7 & 6.5$-$6.9 & $-$18.5 & 28.4$^{+0.1}_{-0.1}$&           9,11,13\\[1ex]
  UDF12-4122-7232 & 03:32:41.22 & $-$27:47:23.2 & 6.8 & 6.5$-$7.0 & $-$17.7 & 29.3$^{+0.3}_{-0.2}$&                  \\[1ex]
  UDF12-3894-7456 & 03:32:38.94 & $-$27:47:45.6 & 6.8 & 6.4$-$7.1 & $-$17.3 & 29.5$^{+0.3}_{-0.3}$&                  \\[1ex]
  UDF12-4290-7174 & 03:32:42.90 & $-$27:47:17.4 & 6.8 & 6.4$-$7.2 & $-$17.3 & 30.2$^{+0.7}_{-0.4}$&                  \\[1ex]
  UDF12-4056-6436 & 03:32:40.56 & $-$27:46:43.6 & 6.8 & 6.7$-$7.0 & $-$18.7 & 28.3$^{+0.1}_{-0.1}$& 1,2,3,4,5,7,9,11,13\\[1ex]
  UDF12-4431-6452 & 03:32:44.31 & $-$27:46:45.2 & 6.8 & 6.6$-$7.0 & $-$18.7 & 28.4$^{+0.1}_{-0.1}$&         1,9,11,13\\[1ex]
  UDF12-3958-6565 & 03:32:39.58 & $-$27:46:56.5 & 6.8 & 6.6$-$7.0 & $-$18.9 & 28.0$^{+0.1}_{-0.1}$&   1,2,3,5,9,11,13\\[1ex]
  UDF12-4037-6560 & 03:32:40.37 & $-$27:46:56.0 & 6.8 & 6.5$-$7.1 & $-$17.5 & 29.7$^{+0.4}_{-0.3}$&              9,13\\[1ex]
  UDF12-4019-6190 & 03:32:40.19 & $-$27:46:19.0 & 6.9 & 6.5$-$7.2 & $-$17.5 & 29.4$^{+0.3}_{-0.2}$&                13\\[1ex]
  UDF12-4422-6337 & 03:32:44.22 & $-$27:46:33.7 & 6.9 & 6.4$-$7.3 & $-$17.1 & 30.2$^{+0.8}_{-0.4}$&                  \\[1ex]
  UDF12-4472-6362 & 03:32:44.72 & $-$27:46:36.2 & 6.9 & 6.5$-$7.1 & $-$18.3 & 28.3$^{+0.1}_{-0.1}$&                13\\[1ex]
  UDF12-4263-6416 & 03:32:42.63 & $-$27:46:41.6 & 6.9 & 6.5$-$7.2 & $-$17.3 & 29.8$^{+0.4}_{-0.3}$&                13\\[1ex]
\hline\hline\end{tabular}
\end{table*}

\setcounter{table}{0}
\begin{table*}
\caption{Candidate $z\geq 6.5$ galaxies in the HUDF continued.}
\begin{tabular}{lccccccc}
\hline
Name & RA(J2000) & Dec(J2000) & $z_{phot}$ &$\Delta z$ & $M_{1500}$ & $H_{160}$ & References\\
\hline
  UDF12-4484-6568 & 03:32:44.84 & $-$27:46:56.8 & 6.9 & 6.5$-$7.1 & $-$17.9 & 29.2$^{+0.4}_{-0.3}$&                  \\[1ex]
  UDF12-3975-7451 & 03:32:39.75 & $-$27:47:45.1 & 6.9 & 6.4$-$7.1 & $-$18.1 & 29.0$^{+0.2}_{-0.2}$&              9,13\\[1ex]
  UDF12-3989-6189 & 03:32:39.89 & $-$27:46:18.9 & 6.9 & 6.5$-$7.1 & $-$18.0 & 29.1$^{+0.2}_{-0.2}$&              1,13\\[1ex]
  UDF12-3729-6175 & 03:32:37.29 & $-$27:46:17.5 & 6.9 & 6.5$-$7.1 & $-$18.2 & 28.5$^{+0.1}_{-0.1}$&                10\\[1ex]
  UDF12-3456-6494 & 03:32:34.56 & $-$27:46:49.4 & 7.0 & 6.6$-$7.3 & $-$17.9 & 28.8$^{+0.1}_{-0.1}$&              9,13\\[1ex]
  UDF12-4068-6498 & 03:32:40.68 & $-$27:46:49.8 & 7.0 & 6.3$-$7.4 & $-$17.9 & 29.2$^{+0.2}_{-0.2}$&                13\\[1ex]
  UDF12-3692-6516 & 03:32:36.92 & $-$27:46:51.6 & 7.0 & 6.6$-$7.4 & $-$17.5 & 29.9$^{+0.5}_{-0.3}$&             10,13\\[1ex]
  UDF12-4071-7347 & 03:32:40.71 & $-$27:47:34.7 & 7.0 & 6.8$-$7.3 & $-$17.8 & 29.7$^{+0.4}_{-0.3}$&              9,13\\[1ex]
  UDF12-4036-8022 & 03:32:40.36 & $-$27:48:02.2 & 7.0 & 6.6$-$7.4 & $-$17.3 & 30.2$^{+0.8}_{-0.5}$&            3,9,13\\[1ex]
  UDF12-3755-6019 & 03:32:37.55 & $-$27:46:01.9 & 7.1 & 6.7$-$7.4 & $-$17.6 & 29.2$^{+0.2}_{-0.2}$&              9,13\\[1ex]
  UDF12-4256-6566 & 03:32:42.56 & $-$27:46:56.6 & 7.1 & 7.0$-$7.2 & $-$20.3 & 26.5$^{+0.1}_{-0.1}$& 1,2,3,4,5,7,9,11,13\\[1ex]
  UDF12-4105-7156 & 03:32:41.05 & $-$27:47:15.6 & 7.1 & 6.8$-$7.3 & $-$19.0 & 28.0$^{+0.1}_{-0.1}$&        1,2,3,5,13\\[1ex]
  UDF12-3853-7519 & 03:32:38.53 & $-$27:47:51.9 & 7.1 & 6.9$-$7.3 & $-$18.0 & 29.3$^{+0.3}_{-0.2}$&            2,4,13\\[1ex]
  UDF12-3825-6566 & 03:32:38.25 & $-$27:46:56.6 & 7.1 & 6.6$-$7.6 & $-$17.5 & 29.4$^{+0.3}_{-0.2}$&                  \\[1ex]
  UDF12-3836-6119 & 03:32:38.36 & $-$27:46:11.9 & 7.1 & 6.8$-$7.3 & $-$18.8 & 28.2$^{+0.1}_{-0.1}$&         3,4,5,9,11\\[1ex]
  UDF12-3709-6441 & 03:32:37.09 & $-$27:46:44.1 & 7.2 & 6.6$-$7.6 & $-$17.0 & 29.8$^{+0.4}_{-0.3}$&                  \\[1ex]
  UDF12-3402-6504 & 03:32:34.02 & $-$27:46:50.4 & 7.2 & 6.9$-$7.3 & $-$18.4 & 28.8$^{+0.1}_{-0.1}$&              9,13\\[1ex]
  UDF12-4384-6311 & 03:32:43.84 & $-$27:46:31.1 & 7.2 & 6.9$-$7.5 & $-$17.9 & 29.5$^{+0.3}_{-0.2}$&                13\\[1ex]
  UDF12-4256-7314 & 03:32:42.56 & $-$27:47:31.4 & 7.2 & 7.1$-$7.4 & $-$19.6 & 27.3$^{+0.1}_{-0.1}$&    1,2,3,4,5,7,9,13\\[1ex]
  UDF12-4035-7468 & 03:32:40.35 & $-$27:47:46.8 & 7.2 & 6.7$-$7.6 & $-$17.5 & 29.8$^{+0.4}_{-0.3}$&                  \\[1ex]
  UDF12-3973-6214 & 03:32:39.73 & $-$27:46:21.4 & 7.3 & 7.1$-$7.4 & $-$18.4 & 28.8$^{+0.1}_{-0.1}$&     1,2,3,4,5,8,9,11\\[1ex] 
  UDF12-3668-8067 & 03:32:36.68 & $-$27:48:06.7 & 7.3 & 7.0$-$7.5 & $-$18.1 & 29.3$^{+0.3}_{-0.2}$&                 9\\[1ex]
  UDF12-3708-8092 & 03:32:37.08 & $-$27:48:09.2 & 7.3 & 6.6$-$7.7 & $-$17.5 & 29.5$^{+0.4}_{-0.3}$&                  \\[1ex]
  UDF12-4242-6243 & 03:32:42.42 & $-$27:46:24.3 & 7.3 & 7.0$-$7.5 & $-$18.4 & 28.5$^{+0.1}_{-0.1}$&             3,5,9\\[1ex] 
  UDF12-3431-7115 & 03:32:34.31 & $-$27:47:11.5 & 7.3 & 7.0$-$7.5 & $-$18.6 & 28.1$^{+0.1}_{-0.1}$&                  \\[1ex]
  UDF12-3868-5477 & 03:32:38.68 & $-$27:45:47.7 & 7.3 & 6.7$-$7.8 & $-$17.1 & 29.9$^{+0.4}_{-0.3}$&                  \\[1ex]
  UDF12-4242-6137 & 03:32:42.42 & $-$27:46:13.7 & 7.3 & 6.9$-$7.7 & $-$18.1 & 28.8$^{+0.2}_{-0.2}$&                  \\[1ex]
  UDF12-4100-7216 & 03:32:41.00 & $-$27:47:21.6 & 7.3 & 6.8$-$7.8 & $-$17.3 &$>30.2$              &                13\\[1ex]
  UDF12-4239-6243 & 03:32:42.39 & $-$27:46:24.3 & 7.3 & 7.0$-$7.5 & $-$18.4 & 28.7$^{+0.2}_{-0.2}$&  3,5,9,13\\[1ex]
  UDF12-4314-6285 & 03:32:43.14 & $-$27:46:28.5 & 7.3 & 6.8$-$7.5 & $-$19.1 & 27.7$^{+0.1}_{-0.1}$&   1,2,3,4,5,6,7,9,11,13\\[1ex]
  UDF12-3313-6545 & 03:32:33.13 & $-$27:46:54.5 & 7.4 & 7.2$-$7.6 & $-$18.6 & 28.6$^{+0.1}_{-0.1}$&       1,4,5,9,11,13\\[1ex]
  UDF12-3885-7540 & 03:32:38.85 & $-$27:47:54.0 & 7.5 & 7.1$-$7.7 & $-$18.2 & 28.9$^{+0.2}_{-0.2}$&                  \\[1ex]
  UDF12-3931-6181 & 03:32:39.31 & $-$27:46:18.1 & 7.5 & 7.3$-$7.8 & $-$18.2 & 29.2$^{+0.2}_{-0.2}$&                13\\[1ex]
  UDF12-4334-6252 & 03:32:43.34 & $-$27:46:25.2 & 7.5 & 7.0$-$8.0 & $-$17.4 & 30.2$^{+0.7}_{-0.4}$&                  \\[1ex]
  UDF12-4308-6242 & 03:32:43.08 & $-$27:46:24.2 & 7.6 & 7.1$-$7.9 & $-$17.7 & 29.2$^{+0.2}_{-0.2}$&         1,9,11,13\\[1ex]
  UDF12-3880-7072 & 03:32:38.80 & $-$27:47:07.2 & 7.7 & 7.5$-$7.8 & $-$20.1 & 26.8$^{+0.1}_{-0.1}$& 1,2,3,4,5,7,9,11,13\\[1ex]
  UDF12-4281-6505 & 03:32:42.81 & $-$27:46:50.5 & 7.7 & 7.2$-$8.0 & $-$17.9 & 28.9$^{+0.2}_{-0.1}$&                  \\[1ex]
  UDF12-4288-6345 & 03:32:42.88 & $-$27:46:34.5 & 7.7 & 7.5$-$7.9 & $-$18.9 & 27.9$^{+0.1}_{-0.1}$&  1,3,4,5,6,8,9,11\\[1ex] 
  UDF12-4470-6443 & 03:32:44.70 & $-$27:46:44.3 & 7.7 & 7.6$-$7.9 & $-$19.7 & 27.4$^{+0.1}_{-0.1}$& 1,2,3,4,5,9,11,13\\[1ex]
  UDF12-4033-8026 & 03:32:40.33 & $-$27:48:02.6 & 7.7 & 7.5$-$8.0 & $-$18.0 & 29.0$^{+0.2}_{-0.2}$&            3,9,13\\[1ex]
  UDF12-3722-8061 & 03:32:37.22 & $-$27:48:06.1 & 7.7 & 7.5$-$7.9 & $-$19.2 & 27.9$^{+0.1}_{-0.1}$&   1,2,3,4,5,9,11,13\\[1ex]
  UDF12-4474-6449 & 03:32:44.74 & $-$27:46:44.9 & 7.8 & 7.5$-$7.9 & $-$18.5 & 28.8$^{+0.1}_{-0.1}$&          1,3,5,13\\[1ex] 
  UDF12-4240-6550 & 03:32:42.40 & $-$27:46:55.0 & 7.8 & 7.5$-$8.1 & $-$17.9 & 29.1$^{+0.2}_{-0.2}$&              9,13\\[1ex]
  UDF12-3939-7040 & 03:32:39.39 & $-$27:47:04.0 & 7.8 & 7.6$-$8.1 & $-$18.4 & 28.5$^{+0.1}_{-0.1}$&                13\\[1ex]
  UDF12-3911-6493 & 03:32:39.11 & $-$27:46:49.3 & 7.9 & 7.5$-$8.2 & $-$18.0 & 29.2$^{+0.2}_{-0.2}$&                 9\\[1ex]
\hline\hline\end{tabular}
\end{table*}

\setcounter{table}{0}
\begin{table*}
\caption{Candidate $z\geq 6.5$ galaxies in the HUDF continued.}
\begin{tabular}{lccccccc}
\hline
Name & RA(J2000) & Dec(J2000) & $z_{phot}$ &$\Delta z$ & $M_{1500}$ & $H_{160}$ & References\\
\hline
  UDF12-3344-6598 & 03:32:33.44 & $-$27:46:59.8 & 7.9 & 7.6$-$8.1 & $-$18.1 & 29.0$^{+0.2}_{-0.2}$&                 9\\[1ex]
  UDF12-3952-7174 & 03:32:39.52 & $-$27:47:17.4 & 7.9 & 7.8$-$8.0 & $-$19.3 & 27.6$^{+0.1}_{-0.1}$&1,2,3,4,5,7,9,11,13\\[1ex]  
  UDF12-4308-6277 & 03:32:43.08 & $-$27:46:27.7 & 8.0 & 7.8$-$8.1 & $-$18.4 & 28.9$^{+0.2}_{-0.1}$&1,3,5,6,7,8,9,11,13\\[1ex]
  UDF12-3780-6001 & 03:32:37.80 & $-$27:46:00.1 & 8.1 & 7.9$-$8.2 & $-$18.9 & 28.3$^{+0.1}_{-0.1}$&   1,3,4,5,6,8,9,13\\[1ex] 
  UDF12-3917-5449 & 03:32:39.17 & $-$27:45:44.9 & 8.1 & 7.8$-$8.5 & $-$17.4 & 29.7$^{+0.4}_{-0.3}$&                \\[1ex]
  UDF12-3762-6011 & 03:32:37.62 & $-$27:46:01.1 & 8.1 & 7.9$-$8.4 & $-$17.9 & 29.4$^{+0.3}_{-0.2}$&                \\[1ex]
  UDF12-3813-5540 & 03:32:38.13 & $-$27:45:54.0 & 8.3 & 8.2$-$8.5 & $-$19.1 & 28.0$^{+0.1}_{-0.1}$&1,3,4,5,6,8,9,11,13\\[1ex]
  UDF12-3763-6015 & 03:32:37.63 & $-$27:46:01.5 & 8.3 & 8.1$-$8.5 & $-$18.7 & 28.4$^{+0.1}_{-0.1}$&      1,3,4,5,6,9,13\\[1ex]
  UDF12-3947-8076 & 03:32:39.47 & $-$27:48:07.6 & 8.6 & 8.4$-$8.8 & $-$18.6 & 28.6$^{+0.1}_{-0.1}$&               9\\[1ex]
  UDF12-3921-6322 & 03:32:39.21 & $-$27:46:32.2 & 8.8 & 8.6$-$9.2 & $-$18.0 & 29.5$^{+0.3}_{-0.2}$&              13\\[1ex]
  UDF12-4344-6547 & 03:32:43.44 & $-$27:46:54.7 & 8.8 & 8.3$-$9.3 & $-$17.6 & 29.7$^{+0.4}_{-0.3}$&              13\\[1ex]
  UDF12-4265-7049 & 03:32:42.65 & $-$27:47:04.9 & 9.5 & 8.8$-$9.9 & $-$18.1 & 29.3$^{+0.2}_{-0.2}$&                \\[1ex]
  UDF12-3954-6285 & 03:32:39.54 & $-$27:46:28.5 & 11.9 & 11.4$-$12.2 & $-$19.7 & 28.9$^{+0.2}_{-0.1}$&           12\\[1ex]
\hline\hline\end{tabular}
\end{table*}

\begin{table*}
\caption{Optical and near-infrared photometry for the candidate $z\geq 6.5$ galaxies in the HUDF. 
In each case the quoted photometry has been corrected to total assuming a point source.
Detections which are less significant than $2\sigma$ are listed as the appropriate $2\sigma$ limit. All candidates are detected at less than $2\sigma$ significance in the 
$B_{435}$, $V_{606}$ and $i_{775}$ filters.}
\begin{tabular}{lccccc}
\hline
Name & $z_{850}$ & $Y_{105}$ & $J_{125}$ & $J_{140}$ & $H_{160}$ \\
\hline
UDF12-3999-6197 & $>30.1$             & 29.4$^{+0.2}_{-0.1}$ & 29.3$^{+0.2}_{-0.2}$  & 29.5$^{+0.3}_{-0.3}$ & 28.8$^{+0.1}_{-0.1}$\\[1ex] 
UDF12-3696-5536 & $>30.1$             & 29.3$^{+0.2}_{-0.2}$ & 29.4$^{+0.3}_{-0.2}$  & 28.8$^{+0.2}_{-0.2}$ & 29.5$^{+0.3}_{-0.2}$\\[1ex] 
UDF12-3677-7536 & 28.6$^{+0.2}_{-0.1}$  & 27.8$^{+0.1}_{-0.1}$ & 27.8$^{+0.1}_{-0.1}$  & 27.9$^{+0.1}_{-0.1}$ & 27.8$^{+0.1}_{-0.1}$\\[1ex] 
UDF12-3897-8116 & $>30.1$             & 29.4$^{+0.2}_{-0.2}$ & 29.3$^{+0.3}_{-0.2}$  & 29.7$^{+0.4}_{-0.3}$ & 29.4$^{+0.3}_{-0.2}$\\[1ex] 
UDF12-4120-6561 & $>30.0$             & 29.6$^{+0.2}_{-0.2}$ & 30.2$^{+0.8}_{-0.4}$  & 30.0$^{+0.5}_{-0.4}$ & $>30.2$\\[1ex] 
UDF12-3515-7257 & $>30.1$              & 29.3$^{+0.2}_{-0.1}$ & 29.6$^{+0.3}_{-0.3}$ & 29.5$^{+0.3}_{-0.3}$ & 29.7$^{+0.4}_{-0.3}$\\[1ex] 
UDF12-3909-6092 & 30.8$^{+0.7}_{-0.4}$  & 29.1$^{+0.1}_{-0.1}$ & 29.3$^{+0.3}_{-0.2}$  & 29.9$^{+0.5}_{-0.3}$ & 29.5$^{+0.3}_{-0.2}$\\[1ex] 
UDF12-3865-6041 & 29.8$^{+0.5}_{-0.4}$  & 29.0$^{+0.1}_{-0.1}$ & 29.0$^{+0.2}_{-0.2}$  & 29.1$^{+0.2}_{-0.2}$ & 29.4$^{+0.3}_{-0.2}$\\[1ex] 
UDF12-3702-5534 & $>30.1$             & 29.6$^{+0.2}_{-0.2}$ & 29.6$^{+0.3}_{-0.3}$  & 29.9$^{+0.5}_{-0.3}$ & 30.2$^{+0.6}_{-0.4}$\\[1ex] 
UDF12-3922-6148 & $>30.1$             & 29.5$^{+0.2}_{-0.2}$ & 29.8$^{+0.4}_{-0.3}$  & 30.1$^{+0.6}_{-0.4}$ & 29.7$^{+0.4}_{-0.3}$\\[1ex] 
UDF12-3736-6245 & $>30.1$             & 29.1$^{+0.1}_{-0.1}$ & 29.1$^{+0.2}_{-0.2}$  & 29.1$^{+0.2}_{-0.2}$ & 29.2$^{+0.2}_{-0.2}$\\[1ex] 
UDF12-4379-6511 & $>30.1$             & 29.0$^{+0.1}_{-0.1}$ & 29.5$^{+0.3}_{-0.3}$  & 29.3$^{+0.3}_{-0.2}$ & 29.4$^{+0.3}_{-0.2}$\\[1ex] 
UDF12-3859-6521 & $>30.1$             & 29.0$^{+0.1}_{-0.1}$ & 29.0$^{+0.2}_{-0.2}$  & 29.6$^{+0.3}_{-0.3}$ & 29.2$^{+0.2}_{-0.2}$\\[1ex] 
UDF12-4202-7074 & $>30.1$             & 29.2$^{+0.1}_{-0.1}$ & 29.2$^{+0.2}_{-0.2}$  & 29.5$^{+0.3}_{-0.2}$ & 29.1$^{+0.2}_{-0.2}$\\[1ex] 
UDF12-3638-7163 & 29.1$^{+0.2}_{-0.2}$  & 28.1$^{+0.1}_{-0.1}$ & 28.1$^{+0.1}_{-0.1} $ & 28.0$^{+0.1}_{-0.1}$ & 28.2$^{+0.1}_{-0.1}$\\[1ex] 
UDF12-4254-6481 & $>30.1$             & 29.6$^{+0.2}_{-0.2}$ & 29.9$^{+0.5}_{-0.3} $ & 29.9$^{+0.5}_{-0.3}$ & 30.2$^{+0.7}_{-0.4}$\\[1ex] 
UDF12-4058-5570 & 29.7$^{+0.5}_{-0.3}$ & 28.9$^{+0.1}_{-0.1}$ & 28.7$^{+0.2}_{-0.1}$   & 29.0$^{+0.2}_{-0.2}$ & 29.1$^{+0.2}_{-0.2}$\\[1ex] 
UDF12-3858-6150 & $>30.1$             & 29.9$^{+0.3}_{-0.2}$ & 29.5$^{+0.3}_{-0.2}$  & 30.0$^{+0.6}_{-0.4}$ & 29.7$^{+0.4}_{-0.3}$\\[1ex] 
UDF12-4186-6322 & 29.9$^{+0.6}_{-0.4}$ & 29.1$^{+0.1}_{-0.1}$ & 28.9$^{+0.2}_{-0.1}$   & 29.4$^{+0.3}_{-0.2}$ & 29.0$^{+0.2}_{-0.2}$\\[1ex] 
UDF12-4144-7041 & $>30.1$            & 29.5$^{+0.2}_{-0.2}$ & 29.3$^{+0.3}_{-0.2}$   & 29.8$^{+0.5}_{-0.3}$ & 29.6$^{+0.4}_{-0.3}$\\[1ex] 
UDF12-4288-6261 & $>30.0$            & 29.4$^{+0.2}_{-0.1}$ & 29.5$^{+0.3}_{-0.2}$   & 29.7$^{+0.4}_{-0.3}$ & 30.1$^{+0.6}_{-0.4}$\\[1ex] 
UDF12-3900-6482 & 30.0$^{+0.7}_{-0.4}$ & 28.7$^{+0.1}_{-0.1}$ & 28.6$^{+0.1}_{-0.1}$   & 28.4$^{+0.1}_{-0.1}$ & 28.2$^{+0.1}_{-0.1}$\\[1ex] 
UDF12-4182-6112 & 30.1$^{+0.8}_{-0.4}$ & 28.9$^{+0.1}_{-0.1}$ & 28.7$^{+0.1}_{-0.1}$   & 28.6$^{+0.1}_{-0.1}$ & 28.5$^{+0.1}_{-0.1}$\\[1ex] 
UDF12-4268-7073 & 29.7$^{+0.5}_{-0.3}$ & 28.7$^{+0.1}_{-0.1}$ & 28.5$^{+0.1}_{-0.1}$   & 28.4$^{+0.1}_{-0.1}$ & 28.5$^{+0.1}_{-0.1}$\\[1ex] 
UDF12-3734-7192 & $>30.1$             & 28.9$^{+0.1}_{-0.1}$ & 28.9$^{+0.2}_{-0.1}$  & 29.1$^{+0.2}_{-0.2}$ & 29.1$^{+0.2}_{-0.2}$\\[1ex] 
UDF12-3968-6066 & $>30.1$             & 29.8$^{+0.3}_{-0.2}$ & 29.4$^{+0.3}_{-0.2}$  & 29.7$^{+0.4}_{-0.3}$ & $>30.2$\\[1ex] 
UDF12-4219-6278 & 28.9$^{+0.2}_{-0.2}$ & 27.7$^{+0.1}_{-0.1}$ & 27.7$^{+0.1}_{-0.1}$   & 27.6$^{+0.1}_{-0.1}$ & 27.7$^{+0.1}_{-0.1}$\\[1ex] 
UDF12-3796-6020 & $>30.1$            & 29.4$^{+0.2}_{-0.1}$ & 29.3$^{+0.3}_{-0.2}$   & 29.4$^{+0.3}_{-0.2}$ & 29.5$^{+0.3}_{-0.2}$\\[1ex] 
UDF12-3675-6447 & 29.9$^{+0.6}_{-0.4}$ & 29.0$^{+0.1}_{-0.1}$ & 28.6$^{+0.1}_{-0.1}$   & 28.7$^{+0.1}_{-0.1}$ & 28.7$^{+0.1}_{-0.1}$\\[1ex] 
UDF12-3744-6513 & 29.2$^{+0.3}_{-0.2}$ & 28.0$^{+0.1}_{-0.1}$ & 27.9$^{+0.1}_{-0.1}$   & 28.1$^{+0.1}_{-0.1}$ & 28.1$^{+0.1}_{-0.1}$\\[1ex] 
UDF12-4160-7045 & 29.6$^{+0.4}_{-0.3}$ & 28.4$^{+0.1}_{-0.1}$ & 28.3$^{+0.1}_{-0.1}$   & 28.3$^{+0.1}_{-0.1}$ & 28.4$^{+0.1}_{-0.1}$\\[1ex] 
UDF12-4122-7232 & $>30.1$            & 29.2$^{+0.1}_{-0.1}$ & 29.2$^{+0.2}_{-0.2}$   & 29.6$^{+0.3}_{-0.3}$ & 29.3$^{+0.3}_{-0.2}$\\[1ex] 
UDF12-3894-7456 & $>30.1$            & 29.5$^{+0.2}_{-0.2}$ & 29.5$^{+0.3}_{-0.3}$   & $>30.2$            & 29.5$^{+0.3}_{-0.3}$\\[1ex] 
UDF12-4290-7174 & $>30.1$            & 29.7$^{+0.2}_{-0.2}$ & 29.4$^{+0.3}_{-0.2}$   & 29.8$^{+0.4}_{-0.3}$ & 30.2$^{+0.7}_{-0.4}$\\[1ex] 
UDF12-4056-6436 & 29.8$^{+0.5}_{-0.4}$ & 28.2$^{+0.1}_{-0.1}$ & 28.2$^{+0.1}_{-0.1}$   & 28.2$^{+0.1}_{-0.1}$ & 28.3$^{+0.1}_{-0.1}$\\[1ex] 
UDF12-4431-6452 & 29.6$^{+0.4}_{-0.3}$ & 28.3$^{+0.1}_{-0.1}$ & 28.2$^{+0.1}_{-0.1}$   & 28.2$^{+0.1}_{-0.1}$ & 28.4$^{+0.1}_{-0.1}$\\[1ex] 
UDF12-3958-6565 & 29.5$^{+0.4}_{-0.3}$ & 28.0$^{+0.1}_{-0.1}$ & 28.0$^{+0.1}_{-0.1}$   & 27.9$^{+0.1}_{-0.1}$ & 28.0$^{+0.1}_{-0.1}$\\[1ex] 
UDF12-4037-6560 & $>30.1$            & 29.5$^{+0.2}_{-0.2}$ & 29.3$^{+0.3}_{-0.2}$   & 29.7$^{+0.4}_{-0.3}$ & 29.7$^{+0.4}_{-0.3}$\\[1ex] 
UDF12-4019-6190 & $>30.1$            & 29.4$^{+0.2}_{-0.1}$ & 29.7$^{+0.4}_{-0.3}$   & 29.5$^{+0.3}_{-0.2}$ & 29.4$^{+0.3}_{-0.2}$\\[1ex] 
UDF12-4422-6337 & $>30.0$            & 29.7$^{+0.3}_{-0.2}$ & $>30.2$              & 29.5$^{+0.4}_{-0.3}$ & 30.2$^{+0.8}_{-0.4}$\\[1ex] 
\hline\hline
\end{tabular}
\end{table*}

\setcounter{table}{1}
\begin{table*}
\caption{Continued.}
\begin{tabular}{lccccc}
\hline
Name & $z_{850}$ & $Y_{105}$ & $J_{125}$ & $J_{140}$ & $H_{160}$ \\[1ex]
\hline
UDF12-4472-6362 & $>30.0$            & 28.7$^{+0.2}_{-0.1}$ & 29.0$^{+0.3}_{-0.2}$ & 28.3$^{+0.2}_{-0.2}$ & 28.3$^{+0.1}_{-0.1}$\\[1ex] 
UDF12-4263-6416 & $>30.0$            & 29.7$^{+0.2}_{-0.2}$ & 29.7$^{+0.4}_{-0.3}$ & 30.0$^{+0.6}_{-0.4}$ & 29.8$^{+0.4}_{-0.3}$\\[1ex] 
UDF12-4484-6568 & $>30.0$            & 28.9$^{+0.2}_{-0.2}$ & 29.5$^{+0.6}_{-0.4}$ & 29.0$^{+0.4}_{-0.3}$ & 29.2$^{+0.4}_{-0.3}$\\[1ex] 
UDF12-3975-7451 & $>30.1$            & 29.0$^{+0.1}_{-0.1}$ & 29.2$^{+0.2}_{-0.2}$ & 28.8$^{+0.2}_{-0.1}$ & 29.0$^{+0.2}_{-0.2}$\\[1ex] 
UDF12-3989-6189 & 30.1$^{+0.8}_{-0.5}$ & 29.0$^{+0.1}_{-0.1}$ & 28.9$^{+0.2}_{-0.1}$ & 29.0$^{+0.2}_{-0.2}$ & 29.1$^{+0.2}_{-0.2}$\\[1ex] 
UDF12-3729-6175 & $>30.1$            & 28.9$^{+0.1}_{-0.1}$ & 28.7$^{+0.1}_{-0.1}$ & 28.8$^{+0.2}_{-0.1}$ & 28.5$^{+0.1}_{-0.1}$\\[1ex] 
UDF12-3456-6494 & $>30.1$            & 29.2$^{+0.1}_{-0.1}$ & 29.1$^{+0.2}_{-0.2}$ & 28.9$^{+0.2}_{-0.2}$ & 28.8$^{+0.1}_{-0.1}$\\[1ex] 
UDF12-4068-6498 & $>30.1$            & 29.3$^{+0.2}_{-0.1}$ & 29.0$^{+0.2}_{-0.2}$ & 29.0$^{+0.2}_{-0.2}$ & 29.2$^{+0.2}_{-0.2}$\\[1ex] 
UDF12-3692-6516 & $>30.1$            & 29.6$^{+0.2}_{-0.2}$ & 29.5$^{+0.3}_{-0.2}$ & 29.3$^{+0.3}_{-0.2}$ & 29.9$^{+0.5}_{-0.3}$\\[1ex] 
UDF12-4071-7347 & $>30.1$            & 29.3$^{+0.1}_{-0.1}$ & 29.6$^{+0.4}_{-0.3}$ & 29.0$^{+0.2}_{-0.2}$ & 29.7$^{+0.4}_{-0.3}$\\[1ex] 
UDF12-4036-8022 & $>30.1$            & 29.7$^{+0.2}_{-0.2}$ & 29.7$^{+0.4}_{-0.3}$ & 29.5$^{+0.3}_{-0.3}$ & 30.2$^{+0.8}_{-0.5}$\\[1ex] 
UDF12-3755-6019 & $>30.1$            & 29.5$^{+0.2}_{-0.2}$ & 29.3$^{+0.2}_{-0.2}$ & 29.6$^{+0.3}_{-0.3}$ & 29.2$^{+0.2}_{-0.2}$\\[1ex] 
UDF12-4256-6566 & 28.9$^{+0.2}_{-0.2}$ & 26.9$^{+0.1}_{-0.1}$ & 26.7$^{+0.1}_{-0.1}$ & 26.6$^{+0.1}_{-0.1}$ & 26.5$^{+0.1}_{-0.1}$\\[1ex] 
UDF12-4105-7156 & 30.0$^{+0.7}_{-0.4}$ & 28.3$^{+0.1}_{-0.1}$ & 28.0$^{+0.1}_{-0.1}$ & 27.9$^{+0.1}_{-0.1}$ & 28.0$^{+0.1}_{-0.1}$\\[1ex] 
UDF12-3853-7519 & $>30.1$            & 29.2$^{+0.1}_{-0.1}$ & 29.0$^{+0.2}_{-0.2}$ & 29.0$^{+0.2}_{-0.2}$ & 29.3$^{+0.3}_{-0.2}$\\[1ex] 
UDF12-3825-6566 & $>30.1$            & 29.7$^{+0.2}_{-0.2}$ & 29.8$^{+0.4}_{-0.3}$ & 29.2$^{+0.2}_{-0.2}$ & 29.4$^{+0.3}_{-0.2}$\\[1ex] 
UDF12-3836-6119 & 29.8$^{+0.6}_{-0.4}$ & 28.5$^{+0.1}_{-0.1}$ & 28.0$^{+0.1}_{-0.1}$ & 28.3$^{+0.1}_{-0.1}$ & 28.2$^{+0.1}_{-0.1}$\\[1ex] 
UDF12-3709-6441 & $>30.1$            & 30.1$^{+0.3}_{-0.3}$ & 30.1$^{+0.6}_{-0.4}$ & 30.1$^{+0.6}_{-0.4}$ & 29.8$^{+0.4}_{-0.3}$\\[1ex] 
UDF12-3402-6504 & $>30.1$            & 28.9$^{+0.1}_{-0.1}$ & 28.4$^{+0.1}_{-0.1}$ & 28.5$^{+0.1}_{-0.1}$ & 28.8$^{+0.1}_{-0.1}$\\[1ex] 
UDF12-4384-6311 & $>30.0$            & 29.5$^{+0.2}_{-0.2}$ & 29.1$^{+0.2}_{-0.2}$ & 29.0$^{+0.2}_{-0.2}$ & 29.5$^{+0.3}_{-0.2}$\\[1ex] 
UDF12-4256-7314 & 29.8$^{+0.6}_{-0.4}$ & 27.7$^{+0.1}_{-0.1}$ & 27.4$^{+0.1}_{-0.1}$ & 27.4$^{+0.1}_{-0.1}$ & 27.3$^{+0.1}_{-0.1}$\\[1ex] 
UDF12-4035-7468 & $>30.0$            & 29.8$^{+0.3}_{-0.2}$ & 29.3$^{+0.3}_{-0.2}$ & 29.7$^{+0.4}_{-0.3}$ & 29.8$^{+0.4}_{-0.3}$\\[1ex] 
UDF12-3973-6214 & $>30.1$            & 28.9$^{+0.1}_{-0.1}$ & 28.4$^{+0.1}_{-0.1}$ & 28.9$^{+0.2}_{-0.1}$ & 28.8$^{+0.1}_{-0.1}$\\[1ex] 
UDF12-3668-8067 & $>30.1$            & 29.2$^{+0.2}_{-0.1}$ & 28.9$^{+0.2}_{-0.2}$ & 28.7$^{+0.2}_{-0.1}$ & 29.3$^{+0.3}_{-0.2}$\\[1ex] 
UDF12-3708-8092 & $>30.1$            & 29.8$^{+0.3}_{-0.2}$ & 29.5$^{+0.4}_{-0.3}$ & 29.4$^{+0.3}_{-0.2}$ & 29.5$^{+0.4}_{-0.3}$\\[1ex] 
UDF12-4242-6243 & $>30.1$            & 29.0$^{+0.1}_{-0.1}$ & 28.6$^{+0.1}_{-0.1}$ & 28.5$^{+0.1}_{-0.1}$ & 28.5$^{+0.1}_{-0.1}$\\[1ex] 
UDF12-3431-7115 & $>30.1$            & 28.8$^{+0.1}_{-0.1}$ & 28.4$^{+0.1}_{-0.1}$ & 28.4$^{+0.1}_{-0.1}$ & 28.1$^{+0.1}_{-0.1}$\\[1ex] 
UDF12-3868-5477 & $>30.1$            & 30.1$^{+0.4}_{-0.3}$ & 30.0$^{+0.5}_{-0.3}$ & 29.8$^{+0.5}_{-0.3}$ & 29.9$^{+0.4}_{-0.3}$\\[1ex] 
UDF12-4242-6137 & $>30.1$            & 29.3$^{+0.2}_{-0.2}$ & 29.1$^{+0.3}_{-0.2}$ & 28.7$^{+0.2}_{-0.2}$ & 28.8$^{+0.2}_{-0.2}$\\[1ex] 
UDF12-4100-7216 & $>30.1$            & 30.1$^{+0.3}_{-0.3}$ & 29.6$^{+0.4}_{-0.3}$ & 29.6$^{+0.4}_{-0.3}$ & $>30.2$\\[1ex] 
UDF12-4239-6243 & 29.6$^{+0.4}_{-0.3}$ & 28.8$^{+0.1}_{-0.1}$ &28.8$^{+0.2}_{-0.1}$  & 28.5$^{+0.1}_{-0.1}$ &28.7$^{+0.1}_{-0.1}$\\[1ex] 
UDF12-4314-6285 & $>30.1$            & 28.6$^{+0.1}_{-0.1}$ &27.9$^{+0.1}_{-0.1}$  & 27.8$^{+0.1}_{-0.1}$ &27.7$^{+0.1}_{-0.1}$\\[1ex] 
UDF12-3313-6545 & $>30.1$            & 28.9$^{+0.1}_{-0.1}$ & 28.5$^{+0.1}_{-0.1}$ & 28.2$^{+0.1}_{-0.1}$ & 28.6$^{+0.1}_{-0.1}$\\[1ex] 
UDF12-3885-7540 & $>30.1$            & 29.4$^{+0.2}_{-0.1}$ & 28.9$^{+0.2}_{-0.2}$ & 28.8$^{+0.2}_{-0.1}$ & 28.9$^{+0.2}_{-0.2}$\\[1ex] 
UDF12-3931-6181 & $>30.1$            & 29.4$^{+0.2}_{-0.2}$ & 28.7$^{+0.1}_{-0.1}$ & 28.8$^{+0.2}_{-0.1}$ & 29.2$^{+0.2}_{-0.2}$\\[1ex] 
UDF12-4334-6252 & $>30.0$            & 30.2$^{+0.5}_{-0.3}$ & 29.4$^{+0.3}_{-0.2}$ & 29.5$^{+0.4}_{-0.3}$ & 30.2$^{+0.7}_{-0.4}$\\[1ex] 
UDF12-4308-6242 & $>30.0$            & 29.9$^{+0.3}_{-0.2}$ & 29.3$^{+0.3}_{-0.2}$ & 29.5$^{+0.4}_{-0.3}$ & 29.2$^{+0.2}_{-0.2}$\\[1ex] 
UDF12-3880-7072 & $>30.1$            & 27.6$^{+0.1}_{-0.1}$ & 27.0$^{+0.1}_{-0.1}$ & 26.9$^{+0.1}_{-0.1}$ & 26.8$^{+0.1}_{-0.1}$\\[1ex] 
UDF12-4281-6505 & $>30.1$            & 30.1$^{+0.3}_{-0.3}$ & 29.2$^{+0.2}_{-0.2}$ & 29.3$^{+0.3}_{-0.2}$ & 28.9$^{+0.2}_{-0.1}$\\[1ex] 
UDF12-4288-6345 & $>30.0$            & 28.9$^{+0.1}_{-0.1}$ & 28.1$^{+0.1}_{-0.1}$ & 28.2$^{+0.1}_{-0.1}$ & 27.9$^{+0.1}_{-0.1}$\\[1ex] 
UDF12-4470-6443 & $>30.0$            & 28.1$^{+0.1}_{-0.1}$ & 27.4$^{+0.1}_{-0.1}$ & 27.4$^{+0.1}_{-0.1}$ & 27.4$^{+0.1}_{-0.1}$\\[1ex] 
UDF12-4033-8026 & $>30.1$            & 29.7$^{+0.2}_{-0.2}$ & 28.9$^{+0.2}_{-0.2}$ & 29.2$^{+0.2}_{-0.2}$ & 29.0$^{+0.2}_{-0.2}$\\[1ex] 
UDF12-3722-8061 & $>30.1$            & 28.6$^{+0.1}_{-0.1}$ & 27.9$^{+0.1}_{-0.1}$ & 27.9$^{+0.1}_{-0.1}$ & 27.9$^{+0.1}_{-0.1}$\\[1ex] 
\hline\hline
\end{tabular}
\end{table*}

\setcounter{table}{1}
\begin{table*}
\caption{Continued.}
\begin{tabular}{lccccc}
\hline
Name & $z_{850}$ & $Y_{105}$ & $J_{125}$ & $J_{140}$ & $H_{160}$ \\[1ex]
\hline
UDF12-4474-6449 & $>30.0$ & 29.4$^{+0.2}_{-0.2}$ & 28.6$^{+0.1}_{-0.1}$ & 28.5$^{+0.1}_{-0.1}$ & 28.8$^{+0.1}_{-0.1}$\\[1ex] 
UDF12-4240-6550 & $>30.1$ & 30.0$^{+0.3}_{-0.2}$ & 29.0$^{+0.2}_{-0.2}$ & 29.3$^{+0.3}_{-0.2}$ & 29.1$^{+0.2}_{-0.2}$\\[1ex] 
UDF12-3939-7040 & $>30.1$ & 29.7$^{+0.2}_{-0.2}$ & 28.7$^{+0.1}_{-0.1}$ & 28.6$^{+0.1}_{-0.1}$ & 28.5$^{+0.1}_{-0.1}$\\[1ex] 
UDF12-3911-6493 & $>30.1$ & 30.0$^{+0.3}_{-0.2}$ & 29.0$^{+0.2}_{-0.2}$ & 29.1$^{+0.2}_{-0.2}$ & 29.2$^{+0.2}_{-0.2}$\\[1ex] 
UDF12-3344-6598 & $>30.1$ & 29.9$^{+0.3}_{-0.2}$ & 28.9$^{+0.2}_{-0.1}$ & 29.0$^{+0.2}_{-0.2}$ & 29.0$^{+0.2}_{-0.2}$\\[1ex] 
UDF12-3952-7174 & $>30.0$ & 28.9$^{+0.1}_{-0.1}$ & 27.9$^{+0.1}_{-0.1}$ & 27.8$^{+0.1}_{-0.1}$ & 27.6$^{+0.1}_{-0.1}$\\[1ex] 
UDF12-4308-6277 & $>30.0$ & 29.8$^{+0.3}_{-0.2}$ & 28.5$^{+0.1}_{-0.1}$ & 29.0$^{+0.2}_{-0.2}$ & 28.9$^{+0.2}_{-0.1}$\\[1ex] 
UDF12-3780-6001 & $>30.1$ & 29.4$^{+0.2}_{-0.2}$ & 28.2$^{+0.1}_{-0.1}$ & 28.3$^{+0.1}_{-0.1}$ & 28.3$^{+0.1}_{-0.1}$\\[1ex] 
UDF12-3917-5449 & $>30.1$ & $>30.6$            & 29.4$^{+0.3}_{-0.2}$ & $>30.1$             & 29.7$^{+0.4}_{-0.3}$\\[1ex] 
UDF12-3762-6011 & $>30.1$ & $>30.7$            & 29.1$^{+0.2}_{-0.2}$ & 29.4$^{+0.3}_{-0.2}$ & 29.4$^{+0.3}_{-0.2}$\\[1ex] 
UDF12-3813-5540 & $>30.1$ & 29.7$^{+0.2}_{-0.2}$ & 28.2$^{+0.1}_{-0.1}$ & 28.1$^{+0.1}_{-0.1}$ & 28.0$^{+0.1}_{-0.1}$\\[1ex] 
UDF12-3763-6015 & $>30.1$ & 30.2$^{+0.4}_{-0.3}$ & 28.5$^{+0.1}_{-0.1}$ & 28.5$^{+0.1}_{-0.1}$ & 28.4$^{+0.1}_{-0.1}$\\[1ex] 
UDF12-3947-8076 & $>30.1$ & 30.7$^{+0.7}_{-0.4}$ & 29.1$^{+0.2}_{-0.2}$ & 28.6$^{+0.1}_{-0.1}$ & 28.6$^{+0.1}_{-0.1}$\\[1ex] 
UDF12-3921-6322 & $>30.1$ & $>30.8$            & 29.5$^{+0.3}_{-0.2}$ & 29.2$^{+0.2}_{-0.2}$ & 29.5$^{+0.3}_{-0.2}$\\[1ex] 
UDF12-4344-6547 & $>30.1$ & $>30.8$            & 29.6$^{+0.4}_{-0.3}$ & 29.8$^{+0.4}_{-0.3}$ & 29.7$^{+0.4}_{-0.3}$\\[1ex] 
UDF12-4265-7049 & $>30.1$ & $>30.8$            & 30.2$^{+0.7}_{-0.4}$ & 29.5$^{+0.3}_{-0.2}$ & 29.3$^{+0.2}_{-0.2}$\\[1ex] 
UDF12-3954-6285 & $>30.1$ & $>30.8$            & $>30.2$            & $>30.2$             & 28.9$^{+0.2}_{-0.1}$\\[1ex]  
\hline\hline
\end{tabular}
\end{table*}

\begin{table*}
\caption{Candidate $z\geq 6.5$ galaxies in HUDF09-1. Column one lists the candidate names and columns two and three list the coordinates. Columns four and five list 
the best-fitting photometric redshift and the corresponding $1\sigma$ uncertainty. Column six lists the total absolute UV magnitude, measured using a top-hat 
filter at 1500\AA\ in the rest-frame of the best-fitting galaxy SED template. Columns 7-10 list the apparent magnitudes in the $z_{850}, Y_{105}, J_{125}\,\&\,H_{160}$ bands, 
which have been corrected to total magnitudes assuming a point source. Detections which are less significant than $2\sigma$ are listed as the appropriate $2\sigma$ limit 
(all candidates are detected at less than $2\sigma$ significance in the $V_{606}$ and $i_{775}$ filters). Column eleven
gives references to previous discoveries of objects: (1)~Bouwens et al. (2011a), (2)~Bouwens et al. (2011a) potential, (3)~Wilkins et al. (2011), (4)~Lorenzoni~et~al.~(2011).}
\begin{tabular}{lcccccccccc}
\hline
Name & RA(J2000) & Dec(J2000) & $z_{phot}$ &$\Delta z$ & $M_{1500}$& $z_{850}$& $Y_{105}$ & $J_{125}$& $H_{160}$ & References\\
\hline
HUDF09-1$\_$40133  &03:32:56.29  &$-$27:40:59.8&6.5  &6.4$-$6.7  &$-18.7$  & 29.1$^{+0.4}_{-0.3}$ & 28.0$^{+0.1}_{-0.1}$ & 28.1$^{+0.2}_{-0.1}$ & 28.7$^{+0.4}_{-0.3}$&\\[1ex]
HUDF09-1$\_$40220  &03:32:56.11  &$-$27:41:20.3&6.5  &6.3$-$6.8  &$-18.3$  & 29.4$^{+0.6}_{-0.4}$ & 28.5$^{+0.2}_{-0.2}$ & 28.5$^{+0.2}_{-0.2}$ & $>29.2$&\\[1ex]
HUDF09-1$\_$30312  &03:33:02.09  &$-$27:41:46.2&6.6  &6.5$-$6.8  &$-19.6$  & 28.4$^{+0.2}_{-0.2}$ & 27.3$^{+0.1}_{-0.1}$ & 27.3$^{+0.1}_{-0.1}$ & 27.4$^{+0.1}_{-0.1}$&\\[1ex]
HUDF09-1$\_$30328  &03:33:02.13  &$-$27:42:00.4&6.7  &6.4$-$7.0  &$-18.6$  & 29.5$^{+0.7}_{-0.4}$ & 28.3$^{+0.2}_{-0.2}$ & 28.3$^{+0.2}_{-0.2}$ & 28.1$^{+0.2}_{-0.2}$&1\\[1ex]
HUDF09-1$\_$40096  &03:32:56.95  &$-$27:40:50.4&6.7  &6.5$-$7.0  &$-18.9$  & 29.1$^{+0.4}_{-0.3}$ & 28.1$^{+0.2}_{-0.1}$ & 27.9$^{+0.1}_{-0.1}$ & 28.2$^{+0.3}_{-0.2}$&1\\[1ex]
HUDF09-1$\_$30126  &03:32:58.99  &$-$27:40:50.0&6.9  &6.8$-$7.2  &$-19.6$  & 29.1$^{+0.4}_{-0.3}$ & 27.5$^{+0.1}_{-0.1}$ & 27.3$^{+0.1}_{-0.1}$ & 27.3$^{+0.1}_{-0.1}$&1\\[1ex]
HUDF09-1$\_$30292  &03:33:02.43  &$-$27:41:31.2&7.0  &6.8$-$7.2  &$-19.5$  & 29.3$^{+0.5}_{-0.3}$ & 27.6$^{+0.1}_{-0.1}$ & 27.4$^{+0.1}_{-0.1}$ & 27.4$^{+0.1}_{-0.1}$&1,3\\[1ex]
HUDF09-1$\_$40030  &03:32:58.74  &$-$27:40:21.5&7.0  &6.7$-$7.2  &$-18.6$  & $>29.6$           & 28.4$^{+0.2}_{-0.2}$ & 28.4$^{+0.2}_{-0.2}$ & 28.6$^{+0.4}_{-0.3}$&1\\[1ex]
HUDF09-1$\_$40181  &03:33:03.81  &$-$27:41:12.3&7.0  &6.7$-$7.2  &$-18.6$  & $>29.6$           & 28.3$^{+0.2}_{-0.2}$ & 28.5$^{+0.2}_{-0.2}$ & 28.6$^{+0.4}_{-0.3}$&1\\[1ex]
HUDF09-1$\_$40131  &03:33:07.29  &$-$27:41:00.1&7.1  &6.7$-$7.5  &$-18.3$  & $>29.6$           & 28.8$^{+0.3}_{-0.2}$ & 28.5$^{+0.2}_{-0.2}$ & $>29.2$&1\\[1ex]
HUDF09-1$\_$30085  &03:32:59.71  &$-$27:40:35.0&7.1  &7.0$-$7.3  &$-19.9$  & 29.4$^{+0.6}_{-0.4}$ & 27.3$^{+0.1}_{-0.1}$ & 27.0$^{+0.1}_{-0.1}$ & 27.1$^{+0.1}_{-0.1}$&3\\[1ex]
HUDF09-1$\_$30185  &03:32:57.82  &$-$27:41:06.5&7.1  &6.8$-$7.5  &$-18.6$  & $>29.6$           & 28.5$^{+0.2}_{-0.2}$ & 28.4$^{+0.2}_{-0.2}$ & 28.3$^{+0.3}_{-0.2}$&1\\[1ex]
HUDF09-1$\_$30239  &03:32:59.59  &$-$27:41:20.6&7.1  &6.9$-$7.3  &$-19.6$  & $>29.6$           & 27.6$^{+0.1}_{-0.1}$ & 27.4$^{+0.1}_{-0.1}$ & 27.4$^{+0.1}_{-0.1}$&3\\[1ex]
HUDF09-1$\_$40037  &03:32:58.53  &$-$27:40:23.5&7.1  &6.9$-$7.3  &$-19.4$  & $>29.6$           & 27.8$^{+0.1}_{-0.1}$ & 27.6$^{+0.1}_{-0.1}$ & 27.7$^{+0.1}_{-0.1}$&1,3\\[1ex]
HUDF09-1$\_$30206  &03:32:56.71  &$-$27:41:07.7&7.4  &7.1$-$7.6  &$-19.9$  & $>29.6$           & 27.6$^{+0.1}_{-0.1}$ & 27.1$^{+0.1}_{-0.1}$ & 27.1$^{+0.1}_{-0.1}$&2,3\\[1ex]
HUDF09-1$\_$40098  &03:32:57.87  &$-$27:40:51.5&7.4  &6.9$-$7.9  &$-18.4$  & $>29.6$           & 29.0$^{+0.4}_{-0.3}$ & 28.5$^{+0.2}_{-0.2}$ & 28.9$^{+0.5}_{-0.4}$&1\\[1ex]
HUDF09-1$\_$30232  &03:32:59.73  &$-$27:41:19.0&7.5  &7.2$-$7.7  &$-19.5$  & $>29.6$           & 28.0$^{+0.1}_{-0.1}$ & 27.6$^{+0.1}_{-0.1}$ & 27.5$^{+0.1}_{-0.1}$&1\\[1ex]
HUDF09-1$\_$40253  &03:32:57.52  &$-$27:41:29.9&7.5  &7.1$-$7.9  &$-18.6$  & $>29.6$           & 28.8$^{+0.4}_{-0.3}$ & 28.4$^{+0.2}_{-0.2}$ & 28.7$^{+0.4}_{-0.3}$&1\\[1ex]
HUDF09-1$\_$30186  &03:32:55.76  &$-$27:41:06.4&7.5  &7.2$-$7.8  &$-19.5$  & $>29.6$           & 28.2$^{+0.2}_{-0.2}$ & 27.6$^{+0.1}_{-0.1}$ & 27.4$^{+0.1}_{-0.1}$&1,3\\[1ex]
HUDF09-1$\_$377  &03:32:59.38  &$-$27:42:01.4&7.6  &7.3$-$7.9    &$-19.1$  & $>29.6$           & 28.5$^{+0.3}_{-0.2}$ & 28.0$^{+0.1}_{-0.1}$ & 27.8$^{+0.2}_{-0.1}$&1\\[1ex]
HUDF09-1$\_$30132  &03:33:04.35  &$-$27:40:51.8&7.6  &7.1$-$8.0  &$-18.5$  & $>29.4$           & 29.1$^{+0.4}_{-0.3}$ & 28.6$^{+0.2}_{-0.2}$ & 28.5$^{+0.3}_{-0.2}$&1\\[1ex]
HUDF09-1$\_$30332  &03:33:02.82  &$-$27:42:02.3&7.6  &7.0$-$8.0  &$-18.4$  & $>29.6$           & 29.2$^{+0.5}_{-0.3}$ & 28.6$^{+0.3}_{-0.2}$ & 28.8$^{+0.4}_{-0.3}$&1\\[1ex]
HUDF09-1$\_$30316  &03:33:00.54  &$-$27:41:46.6&7.7  &7.3$-$8.0  &$-18.8$  & $>29.6$           & 28.9$^{+0.4}_{-0.3}$ & 28.3$^{+0.2}_{-0.2}$ & 28.3$^{+0.3}_{-0.2}$&1,4\\[1ex]
HUDF09-1$\_$50298  &03:33:02.82  &$-$27:42:10.7&7.7  &7.3$-$7.9  &$-19.5$  & $>29.6$           & 28.4$^{+0.2}_{-0.2}$ & 27.6$^{+0.1}_{-0.1}$ & 27.4$^{+0.1}_{-0.1}$&1\\[1ex]
HUDF09-1$\_$30163  &03:32:56.45  &$-$27:41:00.3&7.8  &7.6$-$8.0  &$-19.4$  & $>29.6$           & 28.5$^{+0.2}_{-0.2}$ & 27.7$^{+0.1}_{-0.1}$ & 27.8$^{+0.2}_{-0.1}$&1\\[1ex]
HUDF09-1$\_$30322  &03:33:03.73  &$-$27:41:51.4&7.9  &7.5$-$8.2  &$-18.9$  & $>29.6$           & 29.1$^{+0.4}_{-0.3}$ & 28.1$^{+0.2}_{-0.1}$ & 28.3$^{+0.3}_{-0.2}$&1\\[1ex]
\hline\hline\end{tabular}
\end{table*}

\begin{table*}
\caption{Candidate $z\geq 6.5$ galaxies in HUDF09-2. Column one lists the candidate names and columns two and three list the coordinates. Columns four and five list 
the best-fitting photometric redshift and the corresponding $1\sigma$ uncertainty. Column six lists the total absolute UV magnitude, measured using a top-hat 
filter at 1500\AA\ in the rest-frame of the best-fitting galaxy SED template. Columns 7-10 list the apparent magnitudes in the $z_{850}, Y_{105}, J_{125}\,\&\,H_{160}$ bands, 
which have been corrected to total magnitudes assuming a point source. Detections which are less significant than $2\sigma$ are listed as the appropriate $2\sigma$ limit 
(all candidates are detected at less than $2\sigma$ significance in the $V_{606}$ and $i_{775}$ filters). Column eleven gives 
references to previous discoveries of objects: (1)~Bouwens et al. (2011a), (2)~Bouwens et al. (2011a) potential, (3)~Wilkins et al. (2011),~(4)~McLure~et~al.~(2011).}
\begin{tabular}{lcccccccccc}
\hline
Name & RA(J2000) & Dec(J2000) & $z_{phot}$ &$\Delta z$ & $M_{1500}$& $z_{850}$& $Y_{105}$ & $J_{125}$& $H_{160}$ & References\\
\hline
HUDF09-2$\_$40134  &03:33:05.77  &$-$27:50:55.9&6.5  &6.2$-$6.7  &$-18.3$ & 29.3$^{+0.5}_{-0.4}$ & 28.5$^{+0.2}_{-0.2}$ & 28.6$^{+0.2}_{-0.2}$       &28.8$^{+0.4}_{-0.3}$&\\[1ex]
HUDF09-2$\_$30262  &03:33:00.59  &$-$27:51:49.0&6.6  &6.3$-$6.9  &$-18.3$ & $>29.6$           & 28.4$^{+0.2}_{-0.2}$ & 28.6$^{+0.2}_{-0.2}$       &28.4$^{+0.3}_{-0.2}$&\\[1ex]
HUDF09-2$\_$30274  &03:33:09.14  &$-$27:51:53.1&6.6  &6.4$-$6.8  &$-18.8$ & 29.3$^{+0.5}_{-0.3}$ & 28.0$^{+0.1}_{-0.1}$ & 28.1$^{+0.1}_{-0.1}$       &28.0$^{+0.2}_{-0.1}$&\\[1ex]
HUDF09-2$\_$30224  &03:33:00.80  &$-$27:51:32.0&6.6  &6.4$-$6.9  &$-18.9$ & 29.0$^{+0.4}_{-0.3}$ & 28.1$^{+0.2}_{-0.1}$ & 27.9$^{+0.1}_{-0.1}$       &27.9$^{+0.2}_{-0.1}$&1\\[1ex]
HUDF09-2$\_$40293  &03:33:06.57  &$-$27:51:59.8&6.6  &6.4$-$6.8  &$-18.4$ & $>29.7$           & 28.3$^{+0.2}_{-0.2}$ & 28.4$^{+0.2}_{-0.1}$       &$>29.3$&1\\[1ex]
HUDF09-2$\_$40295  &03:33:01.94  &$-$27:52:03.3&6.7  &6.5$-$6.8  &$-19.6$ & 28.4$^{+0.2}_{-0.1}$ & 27.3$^{+0.1}_{-0.1}$ & 27.3$^{+0.1}_{-0.1}$       &27.2$^{+0.1}_{-0.1}$&4\\[1ex]
HUDF09-2$\_$30085  &03:33:09.77  &$-$27:50:48.5&6.9  &6.7$-$7.1  &$-19.5$ & 29.2$^{+0.5}_{-0.3}$ & 27.7$^{+0.1}_{-0.1}$ & 27.5$^{+0.1}_{-0.1}$       &27.2$^{+0.1}_{-0.1}$&1,3\\[1ex]
HUDF09-2$\_$40120  &03:33:09.64  &$-$27:50:50.8&7.0  &6.9$-$7.1  &$-20.5$ & 28.4$^{+0.2}_{-0.2}$ & 26.5$^{+0.1}_{-0.1}$ & 26.5$^{+0.1}_{-0.1}$       &26.4$^{+0.1}_{-0.1}$&3,4\\[1ex]
HUDF09-2$\_$40282  &03:33:09.14  &$-$27:51:55.5&7.0  &6.8$-$7.1  &$-19.7$ & 29.4$^{+0.6}_{-0.4}$ & 27.2$^{+0.1}_{-0.1}$ & 27.3$^{+0.1}_{-0.1}$       &27.2$^{+0.1}_{-0.1}$&1,3,4\\[1ex]
HUDF09-2$\_$40066  &03:33:07.34  &$-$27:50:41.8&7.0  &6.7$-$7.4  &$-18.1$ & $>29.5$           & 28.7$^{+0.3}_{-0.2}$ & 29.0$^{+0.3}_{-0.2}$       &29.0$^{+0.5}_{-0.3}$&\\[1ex]
HUDF09-2$\_$10162  &03:33:05.40  &$-$27:51:18.9&7.1  &6.9$-$7.3  &$-19.3$ & $>29.6$           & 27.9$^{+0.1}_{-0.1}$ & 27.7$^{+0.1}_{-0.1}$       &27.7$^{+0.1}_{-0.1}$&1,3,4\\[1ex]
HUDF09-2$\_$30145  &03:33:01.19  &$-$27:51:13.4&7.1  &6.9$-$7.3  &$-19.5$ & 29.5$^{+0.7}_{-0.4}$ & 27.8$^{+0.1}_{-0.1}$ & 27.5$^{+0.1}_{-0.1}$       &27.4$^{+0.1}_{-0.1}$&3,4\\[1ex]
HUDF09-2$\_$30132  &03:33:03.81  &$-$27:51:03.4&7.1  &6.6$-$7.6  &$-18.7$ & $>29.5$           & 28.7$^{+0.3}_{-0.2}$ & 28.4$^{+0.2}_{-0.2}$       &28.0$^{+0.2}_{-0.2}$&1\\[1ex]
HUDF09-2$\_$40108  &03:33:09.71  &$-$27:50:48.6&7.2  &6.8$-$7.5  &$-18.6$ & $>29.5$           & 28.7$^{+0.3}_{-0.2}$ & 28.4$^{+0.2}_{-0.1}$       &28.6$^{+0.3}_{-0.2}$&1\\[1ex]
HUDF09-2$\_$40152  &03:33:00.92  &$-$27:51:11.9&7.2  &6.8$-$7.6  &$-18.2$ & $>29.6$           & 28.8$^{+0.3}_{-0.3}$ & 28.8$^{+0.3}_{-0.2}$       &28.8$^{+0.4}_{-0.3}$&1\\[1ex]
HUDF09-2$\_$40020  &03:33:07.09  &$-$27:50:21.8&7.2  &6.8$-$7.6  &$-18.6$ & $>29.5$           & 28.8$^{+0.3}_{-0.2}$ & 28.3$^{+0.2}_{-0.1}$       &28.8$^{+0.4}_{-0.3}$&2\\[1ex]
HUDF09-2$\_$10189  &03:33:06.39  &$-$27:51:24.8&7.7  &7.5$-$8.0  &$-18.9$ & $>29.6$           & 28.9$^{+0.3}_{-0.3}$ & 28.1$^{+0.1}_{-0.1}$       &28.4$^{+0.3}_{-0.2}$&1\\[1ex]
HUDF09-2$\_$30170  &03:33:03.78  &$-$27:51:20.4&7.7  &7.6$-$7.9  &$-20.6$ & $>29.6$           & 27.2$^{+0.1}_{-0.1}$ & 26.5$^{+0.1}_{-0.1}$       &26.3$^{+0.1}_{-0.1}$&1,3,4\\[1ex]
HUDF09-2$\_$50096  &03:33:04.64  &$-$27:50:53.0&7.8  &7.6$-$7.9  &$-19.8$ & $>29.5$           & 28.0$^{+0.2}_{-0.1}$ & 27.3$^{+0.1}_{-0.1}$       &27.3$^{+0.1}_{-0.1}$&1,4 \\[1ex]
HUDF09-2$\_$10164  &03:33:03.76  &$-$27:51:19.7&7.8  &7.7$-$7.9  &$-20.3$ & $>29.6$           & 27.6$^{+0.1}_{-0.1}$ & 26.8$^{+0.1}_{-0.1}$       &26.7$^{+0.1}_{-0.1}$&1,4 \\[1ex]
HUDF09-2$\_$10026  &03:33:06.97  &$-$27:50:27.9&8.4  &7.9$-$8.7  &$-18.8$ & $>29.5$           &  $>29.5$          & 28.5$^{+0.2}_{-0.2}$       &28.5$^{+0.3}_{-0.2}$&1\\[1ex]
HUDF09-2$\_$50121  &03:33:03.39  &$-$27:51:00.4&8.4  &7.9$-$8.7  &$-19.0$ & $>29.5$           &  $>29.5$          & 28.3$^{+0.2}_{-0.1}$       &28.2$^{+0.2}_{-0.2}$&1\\[1ex]
HUDF09-2$\_$50104  &03:33:07.58  &$-$27:50:55.1&9.0  &8.6$-$9.2  &$-19.9$ & $>29.6$           &  $>29.6$          & 27.9$^{+0.1}_{-0.1}$        &27.4$^{+0.1}_{-0.1}$&1,4\\[1ex]
HUDF09-2$\_$247  &03:33:04.24  &$-$27:52:09.4&9.4   &9.1$-$9.6   &$-19.5$ & $>29.7$           &  $>29.6$          & 28.8$^{+0.2}_{-0.2}$       &27.9$^{+0.2}_{-0.1}$&1\\[1ex]
\hline\hline\end{tabular}
\end{table*}

\begin{table*}
\caption{Candidate $z\geq 6.5$ galaxies in ERS. Column one lists the candidate names and columns two and three list the coordinates. Columns four and five list 
the best-fitting photometric redshift and the corresponding $1\sigma$ uncertainty. Column six lists the total absolute UV magnitude, measured using a top-hat 
filter at 1500\AA\ in the rest-frame of the best-fitting galaxy SED template. Columns 7-10 list the apparent magnitudes in the $z_{850}, Y_{098}, J_{125}\,\&\,H_{160}$ bands, 
which have been corrected to total magnitudes assuming a point source. Detections which are less significant than $2\sigma$ are listed as the appropriate $2\sigma$ limit 
(all candidates are detected at less than $2\sigma$ significance in the $B_{435}, V_{606}$ and $i_{775}$ filters). Column eleven
gives references to previous discoveries of objects: (1)~Bouwens et al. (2011a), (2)~Bouwens et al. (2011a) potential, (3)~Wilkins et al. (2011), (4)~Lorenzoni et al. (2011), (5)~McLure et al. (2011).}
\begin{tabular}{lcccccccccc}
\hline
Name & RA(J2000) & Dec(J2000) & $z_{phot}$ &$\Delta z$ & $M_{1500}$& $z_{850}$& $Y_{098}$ & $J_{125}$& $H_{160}$ & References\\
\hline
ERS$\_$30059  &03:32:21.81  &$-$27:44:38.7&6.5  &6.2$-$6.7  &$-19.8$ & 28.0$^{+0.4}_{-0.3}$ & 27.1$^{+0.2}_{-0.1}$ & 27.1$^{+0.1}_{-0.1}$           &26.8$^{+0.1}_{-0.1}$&\\[1ex]
ERS$\_$50656  &03:32:27.01  &$-$27:41:42.9&6.5  &6.0$-$6.8  &$-19.4$ & 28.0$^{+0.4}_{-0.3}$ & 27.4$^{+0.2}_{-0.2}$ & 27.3$^{+0.2}_{-0.1}$           &27.6$^{+0.3}_{-0.2}$&2\\[1ex]
ERS$\_$30220  &03:32:41.39  &$-$27:43:16.9&6.6  &6.4$-$6.8  &$-20.3$ & 27.6$^{+0.3}_{-0.2}$ & 26.6$^{+0.1}_{-0.1}$ & 26.6$^{+0.1}_{-0.1}$           &26.3$^{+0.1}_{-0.1}$&2\\[1ex]
ERS$\_$30412  &03:32:07.86  &$-$27:42:17.8&6.6  &6.5$-$6.9  &$-20.4$ & 27.3$^{+0.3}_{-0.2}$ & 26.7$^{+0.1}_{-0.1}$ & 26.4$^{+0.1}_{-0.1}$           &26.4$^{+0.1}_{-0.1}$&5\\[1ex]
ERS$\_$30089  &03:32:06.83  &$-$27:44:22.2&6.7  &6.5$-$6.8  &$-20.2$ & 27.9$^{+0.4}_{-0.3}$ & 26.6$^{+0.1}_{-0.1}$ & 26.6$^{+0.1}_{-0.1}$           &26.8$^{+0.2}_{-0.1}$&1,5\\[1ex]
ERS$\_$30172  &03:32:20.24  &$-$27:43:34.3&6.7  &6.4$-$7.0  &$-19.6$ & 28.2$^{+0.6}_{-0.4}$ & 27.4$^{+0.2}_{-0.2}$ & 27.2$^{+0.1}_{-0.1}$            &27.0$^{+0.2}_{-0.2}$&2,5\\[1ex]
ERS$\_$30562  &03:32:22.52  &$-$27:41:17.4&6.7  &6.5$-$6.9  &$-19.8$ & 28.3$^{+0.7}_{-0.4}$ & 27.1$^{+0.2}_{-0.2}$ & 27.1$^{+0.2}_{-0.1}$           &27.0$^{+0.2}_{-0.2}$&1\\[1ex]
ERS$\_$30457  &03:32:29.54  &$-$27:42:04.5&6.9  &6.7$-$7.0  &$-20.1$ & $>28.4$           & 27.0$^{+0.2}_{-0.2}$ & 26.8$^{+0.1}_{-0.1}$           &26.8$^{+0.2}_{-0.2}$&1,3,5\\[1ex]
ERS$\_$30645  &03:32:16.00  &$-$27:41:59.1&7.0  &6.7$-$7.2  &$-19.9$ & $>28.3$           & 27.5$^{+0.3}_{-0.2}$ & 27.1$^{+0.1}_{-0.1}$           &27.0$^{+0.2}_{-0.2}$&1,5\\[1ex]
ERS$\_$30426  &03:32:24.09  &$-$27:42:13.8&7.0  &6.8$-$7.2  &$-20.2$ & 28.4$^{+0.7}_{-0.4}$ & 27.2$^{+0.2}_{-0.2}$ & 26.8$^{+0.1}_{-0.1}$           &26.6$^{+0.2}_{-0.1}$&1,3,5\\[1ex]
ERS$\_$30618  &03:32:16.19  &$-$27:41:49.8&7.0  &6.7$-$7.2  &$-19.6$ & $>28.3$           & 27.7$^{+0.4}_{-0.3}$ & 27.3$^{+0.2}_{-0.2}$           &27.3$^{+0.3}_{-0.2}$&1,5\\[1ex]
ERS$\_$10317  &03:32:03.21  &$-$27:42:58.1&7.3  &7.0$-$7.6  &$-19.5$ & $>28.5$           & $>28.2$           & 27.4$^{+0.2}_{-0.2}$            &27.9$^{+0.6}_{-0.4}$&\\[1ex]
ERS$\_$30100  &03:32:11.16  &$-$27:44:16.9&7.4  &7.2$-$7.5  &$-20.1$ & $>28.3$           & 28.0$^{+0.4}_{-0.3}$ & 26.9$^{+0.1}_{-0.1}$           &26.9$^{+0.2}_{-0.1}$&1\\[1ex]
ERS$\_$50207  &03:32:35.46  &$-$27:43:23.5&7.4  &7.0$-$7.7  &$-19.5$ & $>28.5$           & $>28.2$           & 27.5$^{+0.2}_{-0.2}$           &27.6$^{+0.4}_{-0.3}$&\\[1ex]
ERS$\_$50461  &03:32:25.16  &$-$27:41:57.4&7.5  &7.3$-$7.8  &$-19.8$ & $>28.5$           & $>28.3$           & 27.2$^{+0.2}_{-0.2}$           &27.2$^{+0.3}_{-0.2}$&1\\[1ex]
ERS$\_$10237  &03:32:23.37  &$-$27:43:26.5&8.0  &7.6$-$8.3  &$-19.8$ & $>28.5$           & $>28.2$           & 27.1$^{+0.2}_{-0.1}$           &27.7$^{+0.5}_{-0.3}$&5\\[1ex]
ERS$\_$50613  &03:32:35.44  &$-$27:41:32.7&8.4  &7.5$-$8.7  &$-20.0$ & $>28.4$           & $>28.5$           & 27.2$^{+0.2}_{-0.1}$           &27.2$^{+0.2}_{-0.2}$&1,5\\[1ex]
ERS$\_$50150  &03:32:02.99  &$-$27:43:51.9&8.5  &7.7$-$8.8  &$-20.3$ & $>28.4$           & $>28.2$           & 27.0$^{+0.1}_{-0.1}$           &26.9$^{+0.2}_{-0.2}$&1,4,5\\[1ex]
\hline\hline\end{tabular}
\end{table*}

\begin{table*}
\caption{Candidate $z\geq 6.5$ galaxies in CANDELS GS-DEEP. Column one lists the candidate names and columns two and three list the coordinates. Columns four and five list 
the best-fitting photometric redshift and the corresponding $1\sigma$ uncertainty. Column six lists the total absolute UV magnitude, measured using a top-hat 
filter at 1500\AA\ in the rest-frame of the best-fitting galaxy SED template.  Columns 7-10 list the apparent magnitudes in the $z_{850}, Y_{105}, J_{125}\,\&\,H_{160}$ bands, 
which have been corrected to total magnitudes assuming a point source. Detections which are less significant than $2\sigma$ are listed as the appropriate $2\sigma$ limit 
(all candidates are detected at less than $2\sigma$ significance in the $B_{435}, V_{606}$ and $i_{775}$ filters). Those objects which 
have significantly deeper $z_{850}$ limits lie within the footprint of the original HUDF ACS imaging. Column eleven
gives references to previous discoveries of objects: (1)~Oesch et al. (2012), (2)~Yan et al. (2012), (3)~Grazian et al. (2012), (4)~Lorenzoni et al. (2013).} 
\begin{tabular}{lcccccccccc}
\hline
Name & RA(J2000) & Dec(J2000) & $z_{phot}$ &$\Delta z$ & $M_{1500}$& $z_{850}$& $Y_{105}$ & $J_{125}$& $H_{160}$ & References\\
\hline
CGSD$\_$130052 &03:32:37.23  &$-$27:45:38.4&6.5  &6.4$-$6.6  &$-19.7$ & 28.1$^{+0.1}_{-0.1}$ & 27.2$^{+0.1}_{-0.1}$ & 27.1$^{+0.1}_{-0.1}$                   &27.2$^{+0.2}_{-0.2}$&4\\[1ex]
CGSD$\_$30322  &03:32:22.68  &$-$27:46:09.1&6.6  &6.1$-$7.0  &$-18.9$ & $>28.6$           & 27.9$^{+0.3}_{-0.2}$ & 27.9$^{+0.3}_{-0.2}$                   &28.0$^{+0.5}_{-0.3}$&\\[1ex]
CGSD$\_$30157  &03:32:37.28  &$-$27:48:54.6&6.7  &6.4$-$7.0  &$-21.2$ & 27.6$^{+0.1}_{-0.1}$ & 26.0$^{+0.1}_{-0.1}$ & 25.5$^{+0.1}_{-0.1}$                   &25.5$^{+0.1}_{-0.1}$&\\[1ex]
CGSD$\_$30134  &03:32:14.24  &$-$27:48:55.3&6.8  &6.5$-$7.1  &$-19.7$ &$>28.6$            & 27.2$^{+0.2}_{-0.2}$ & 27.3$^{+0.2}_{-0.1}$                    &27.1$^{+0.2}_{-0.2}$&\\[1ex]
CGSD$\_$30237  &03:32:28.35  &$-$27:47:34.6&6.8  &6.6$-$7.1  &$-19.2$ &$>29.8$            & 27.9$^{+0.3}_{-0.2}$ & 27.7$^{+0.2}_{-0.2}$                   &27.8$^{+0.4}_{-0.3}$&\\[1ex]
CGSD$\_$30261  &03:32:40.68  &$-$27:45:11.6&6.9  &6.5$-$7.3  &$-19.5$ &$>28.5$            & 27.6$^{+0.2}_{-0.2}$ & 27.5$^{+0.2}_{-0.2}$                   &27.2$^{+0.2}_{-0.2}$&2\\[1ex]
CGSD$\_$30275  &03:32:19.94  &$-$27:47:10.6&6.9  &6.7$-$7.2  &$-19.7$ &$>28.4$            & 27.2$^{+0.2}_{-0.1}$ & 27.3$^{+0.2}_{-0.1}$                    &27.2$^{+0.2}_{-0.2}$&3,4\\[1ex]
CGSD$\_$30239  &03:32:42.56  &$-$27:47:31.4&7.0  &6.7$-$7.2  &$-19.5$ &$>30.2$            & 27.6$^{+0.2}_{-0.2}$ & 27.4$^{+0.2}_{-0.2}$                   &27.3$^{+0.2}_{-0.2}$&\\[1ex]
CGSD$\_$30336  &03:32:25.22  &$-$27:46:26.7&7.3  &7.1$-$7.5  &$-20.2$ &$>28.7$            & 27.1$^{+0.1}_{-0.1}$ & 26.7$^{+0.1}_{-0.1}$                    &26.8$^{+0.1}_{-0.1}$&\\[1ex]
CGSD$\_$30389  &03:32:40.69  &$-$27:44:16.7&7.3  &7.1$-$7.5  &$-20.2$ &$>28.5$            & 27.2$^{+0.1}_{-0.1}$ & 26.8$^{+0.1}_{-0.1}$                    &26.8$^{+0.1}_{-0.1}$&4\\[1ex]
CGSD$\_$30146  &03:32:23.77  &$-$27:49:13.6&7.3  &7.1$-$7.6  &$-20.1$ &$>28.6$            & 27.2$^{+0.2}_{-0.2}$ & 26.9$^{+0.1}_{-0.1}$                    &27.0$^{+0.2}_{-0.2}$&3\\[1ex]
CGSD$\_$30159  &03:32:24.33  &$-$27:49:15.0&7.3  &7.0$-$7.6  &$-19.8$ &$>28.6$            & 27.5$^{+0.3}_{-0.2}$ & 27.2$^{+0.2}_{-0.1}$                   &27.1$^{+0.2}_{-0.2}$&1,3\\[1ex]
CGSD$\_$30048  &03:32:46.89  &$-$27:50:07.5&7.4  &7.2$-$7.7  &$-20.4$ &$>28.4$            & 27.1$^{+0.1}_{-0.1}$ & 26.7$^{+0.1}_{-0.1}$                   &26.4$^{+0.1}_{-0.1}$&1\\[1ex]
CGSD$\_$30284  &03:32:27.91  &$-$27:45:42.8&7.4  &7.1$-$7.7  &$-19.8$ &$>28.8$            & 27.7$^{+0.2}_{-0.2}$ & 27.2$^{+0.1}_{-0.1}$                   &27.6$^{+0.3}_{-0.2}$&4\\[1ex]
CGSD$\_$30388  &03:32:40.26  &$-$27:44:09.9&7.5  &7.1$-$7.8  &$-19.7$ &$>28.6$            & 27.8$^{+0.2}_{-0.2}$ & 27.4$^{+0.2}_{-0.2}$                    &27.3$^{+0.3}_{-0.2}$&1,4\\[1ex]
CGSD$\_$40222  &03:32:32.03  &$-$27:45:37.1&7.6  &7.4$-$7.8  &$-20.6$ &$>28.6$            & 27.1$^{+0.1}_{-0.1}$ & 26.5$^{+0.1}_{-0.1}$                    &26.3$^{+0.1}_{-0.1}$&1\\[1ex]
CGSD$\_$130048 &03:32:44.02  &$-$27:47:27.3&7.7  &7.5$-$7.9  &$-20.0$ &$>29.7$            & 27.6$^{+0.2}_{-0.2}$ & 27.1$^{+0.1}_{-0.1}$                    &26.9$^{+0.2}_{-0.1}$&4\\[1ex]
CGSD$\_$30414  &03:32:47.95  &$-$27:44:50.4&7.7  &7.4$-$8.0  &$-19.8$ &$>28.3$            & 28.1$^{+0.3}_{-0.2}$ & 27.4$^{+0.2}_{-0.2}$                    &27.0$^{+0.2}_{-0.2}$&2\\[1ex]
CGSD$\_$30014  &03:32:21.43  &$-$27:52:21.5&7.8  &7.4$-$8.3  &$-19.5$ &$>28.4$            & 28.4$^{+0.7}_{-0.4}$ & 27.5$^{+0.2}_{-0.2}$                   &27.9$^{+0.5}_{-0.3}$&\\[1ex]
CGSD$\_$50050  &03:32:20.97  &$-$27:51:37.1&7.9  &7.6$-$8.1  &$-20.0$ &$>28.4$            & 28.0$^{+0.2}_{-0.2}$ & 27.1$^{+0.1}_{-0.1}$                    &26.9$^{+0.2}_{-0.1}$&1,2\\[1ex]
CGSD$\_$30337  &03:32:14.13  &$-$27:48:28.9&7.9  &7.6$-$8.3  &$-20.2$ &$>28.5$            & 27.9$^{+0.4}_{-0.3}$ & 26.9$^{+0.1}_{-0.1}$                   &27.1$^{+0.2}_{-0.2}$&2,4\\[1ex]
CGSD$\_$10080  &03:32:14.21  &$-$27:50:06.9&8.0  &7.4$-$8.3  &$-19.3$ &$>28.5$            & $>28.4$           & 27.7$^{+0.2}_{-0.2}$                   &28.4$^{+0.8}_{-0.4}$&\\[1ex]
CGSD$\_$50145  &03:32:49.94  &$-$27:48:18.1&8.0  &7.9$-$8.1  &$-21.4$ &$>28.5$            & 27.0$^{+0.1}_{-0.1}$ & 25.8$^{+0.1}_{-0.1}$                   &25.7$^{+0.1}_{-0.1}$&1,2\\[1ex]  
CGSD$\_$50276  &03:32:27.79  &$-$27:45:14.1&8.2  &7.8$-$8.5  &$-19.7$ &$>28.6$            & 28.8$^{+0.6}_{-0.4}$ & 27.4$^{+0.2}_{-0.2}$                   &27.6$^{+0.3}_{-0.3}$&1\\[1ex]
CGSD$\_$10023  &03:32:16.91  &$-$27:52:01.9&8.2  &7.8$-$8.6  &$-19.5$ &$>28.5$            & $>28.5$           & 27.6$^{+0.2}_{-0.2}$                  &28.1$^{+0.6}_{-0.4}$&2\\[1ex]
CGSD$\_$50001  &03:32:18.19  &$-$27:52:45.6&8.4  &8.0$-$8.6  &$-20.2$ &$>28.5$            & $>28.6$           & 27.1$^{+0.2}_{-0.1}$                  &27.0$^{+0.2}_{-0.2}$&1,2\\[1ex]
CGSD$\_$50067  &03:32:35.00  &$-$27:49:21.6&8.9  &8.7$-$9.1  &$-20.3$ &$>28.5$            & $>28.9$           & 27.3$^{+0.2}_{-0.1}$                   &27.0$^{+0.2}_{-0.1}$&1,2\\[1ex]
\hline\hline\end{tabular}
\end{table*}

\begin{table*}
\caption{Candidate $z\geq 6.5$ galaxies in CANDELS GS-WIDE. Column one lists the candidate names and columns two and three list the coordinates. Columns four and five list 
the best-fitting photometric redshift and the corresponding $1\sigma$ uncertainty. Column six lists the total absolute UV magnitude, measured using a top-hat 
filter at 1500\AA\ in the rest-frame of the best-fitting galaxy SED template.  Columns 7-10 list the apparent magnitudes in the $z_{850}, Y_{105}, J_{125}\,\&\,H_{160}$ bands, 
which have been corrected to total magnitudes assuming a point source. Detections which are less significant than $2\sigma$ are listed as the appropriate $2\sigma$ limit 
(both candidates are detected at less than $2\sigma$ significance in the $B_{435}, V_{606}$ and $i_{775}$ filters). Column eleven
gives references to previous discoveries of objects.}
\begin{tabular}{lcccccccccc}
\hline
Name & RA(J2000) & Dec(J2000) & $z_{phot}$ &$\Delta z$ & $M_{1500}$& $z_{850}$& $Y_{105}$ & $J_{125}$& $H_{160}$ & References\\
\hline
CGSW$\_$30099  &03:32:57.39  &$-$27:53:21.8&6.5  &6.4$-$6.6  &$-20.1$  & 27.6$^{+0.1}_{-0.1}$ & 26.6$^{+0.2}_{-0.2}$ & 26.7$^{+0.2}_{-0.2}$  &26.7$^{+0.3}_{-0.2}$&\\[1ex]
CGSW$\_$30210  &03:32:57.62  &$-$27:52:37.4&6.5  &6.4$-$6.7  &$-20.0$  & 27.7$^{+0.1}_{-0.1}$ & 26.6$^{+0.2}_{-0.2}$ & 27.0$^{+0.3}_{-0.2}$  &26.6$^{+0.3}_{-0.2}$&\\[1ex]
\hline\hline\end{tabular}
\end{table*}

\begin{table*}
\caption{Candidate $z\geq 6.5$ galaxies in CANDELS UDS. Column one lists the candidate names and columns two and three list the coordinates. Columns four and five list 
the best-fitting photometric redshift and the corresponding $1\sigma$ uncertainty. Column six lists the total absolute UV magnitude, measured using a top-hat 
filter at 1500\AA\ in the rest-frame of the best-fitting galaxy SED template.  Columns 7-9 list the apparent magnitudes in the $i_{814}, J_{125}\,\&\,H_{160}$ bands, 
which have been corrected to total magnitudes assuming a point source. Detections which are less significant than $2\sigma$ are listed as the appropriate $2\sigma$ limit 
(all candidates are detected at less than $2\sigma$ significance in the $V_{606}$ filter). Column ten
gives references to previous discoveries of objects: (1)~Ouchi et al. (2009), (2)~Grazian et al. (2012). We note that CUDS$\_$20114 is an extended, multi-component, 
object (Ouchi et al. 2009), and that the photometry reported here is for the central component only.}
\begin{tabular}{lccccccccc}
\hline
Name & RA(J2000) & Dec(J2000) & $z_{phot}$ &$\Delta z$ & $M_{1500}$ & $i_{814}$ & $J_{125}$ & $H_{160}$ & References\\
\hline
CUDS$\_$20144  &02:16:54.99  &$-$05:09:11.9&6.5  &6.4$-$6.6  &$-20.4$  & $>28.7$ & 26.3$^{+0.2}_{-0.1}$  &26.3$^{+0.2}_{-0.1}$&2\\[1ex]
CUDS$\_$20114  &02:17:57.61  &$-$05:08:44.9&6.6  &6.5$-$6.7  &$-20.8$  & $>28.2$ & 26.0$^{+0.1}_{-0.1}$  &26.1$^{+0.1}_{-0.1}$&1,2\\[1ex]
CUDS$\_$20253  &02:17:11.61  &$-$05:10:33.7&6.7  &6.5$-$7.0  &$-20.5$  & $>28.6$ & 26.3$^{+0.2}_{-0.1}$  &26.2$^{+0.2}_{-0.1}$&2\\[1ex]
CUDS$\_$20482  &02:17:39.70  &$-$05:13:50.6&6.8  &6.5$-$7.0  &$-20.7$  & $>28.6$ & 26.1$^{+0.1}_{-0.1}$  &26.2$^{+0.1}_{-0.1}$&2\\[1ex]
CUDS$\_$20450  &02:17:15.43  &$-$05:13:23.7&7.1  &6.9$-$7.4  &$-20.3$  & $>28.7$ & 26.8$^{+0.2}_{-0.2}$  &26.5$^{+0.2}_{-0.2}$&\\[1ex]
CUDS$\_$20398  &02:16:56.13  &$-$05:12:36.1&7.2  &7.1$-$7.3  &$-21.1$  & $>28.7$ & 25.9$^{+0.1}_{-0.1}$  &25.9$^{+0.1}_{-0.1}$&2\\[1ex]
CUDS$\_$20615   &02:17:41.33  &$-$05:15:33.4&7.4  &7.1$-$7.7  &$-21.0$  & $>28.3$ & 26.1$^{+0.1}_{-0.1}$   &25.8$^{+0.1}_{-0.1}$&2\\[1ex]
\hline\hline\end{tabular}
\end{table*}

\begin{table*}
\caption{Candidate $z\geq 6.5$ galaxies in BoRG. Column one lists the candidate names and columns two and three list the coordinates. Columns four and five list 
the best-fitting photometric redshift and the corresponding $1\sigma$ uncertainty. Column six lists the total absolute UV magnitude, measured using a top-hat 
filter at 1500\AA\ in the rest-frame of the best-fitting galaxy SED template.  Columns 7-9 list the apparent magnitudes in the $Y_{098}, J_{125}\,\&\,H_{160}$ bands, 
which have been corrected to total magnitudes assuming a point source. Detections which are less significant than $2\sigma$ are listed as the appropriate $2\sigma$ limit 
(all candidates are detected at less than $2\sigma$ significance in the $V_{606}$ or $V_{600}$ filters). Column ten
gives references to previous discoveries of objects: (1)~Bradley et al. (2012).}
\begin{tabular}{lccccccccc}
\hline
Name & RA(J2000) & Dec(J2000) & $z_{phot}$ &$\Delta z$ & $M_{1500}$ & $Y_{098}$ &$J_{125}$ & $H_{160}$ & References\\
\hline
BoRG$\_$54  &08:35:13.61  &$+$24:55:36.2&7.6  &7.4$-$7.8  &$-20.7$  & $>28.0$  & 26.3$^{+0.1}_{-0.1}$  &26.5$^{+0.3}_{-0.2}$&1\\[1ex]
BoRG$\_$120  &22:02:46.32  &$+$18:51:29.5&7.7  &7.3$-$8.0  &$-20.4$ & $>28.8$  & 26.6$^{+0.2}_{-0.1}$  &26.7$^{+0.2}_{-0.2}$&1\\[1ex]
BoRG$\_$118  &22:02:50.37  &$+$18:50:21.0&8.3  &7.5$-$8.8  &$-20.4$ & $>28.8$  & 26.9$^{+0.2}_{-0.2}$  &26.9$^{+0.3}_{-0.2}$&\\[1ex]
BoRG$\_$51  &04:39:46.95  &$-$52:43:55.2&8.4  &7.6$-$8.8  &$-21.4$  & $>28.6$  & 26.0$^{+0.1}_{-0.1}$  &25.9$^{+0.1}_{-0.1}$&1\\[1ex]
\hline\hline\end{tabular}
\end{table*}

\end{appendix}

\end{document}